\documentclass[12pt]{article}
\usepackage{latexsym}
\usepackage{epsfig,amssymb,euscript,slashed}
\usepackage{amsmath}
\usepackage{array,calc,epsfig}
\usepackage[linktocpage]{hyperref}
\usepackage{pstricks}
\usepackage{bbm}
\usepackage{fancybox}

\def\be{\begin{equation}}
\def\ee{\end{equation}}
\def\bseq{\begin{subequations}}
\def\eseq{\end{subequations}}

\def\bea{\begin{eqnarray}}
\def\eea{\end{eqnarray}}
\newcommand\bbone{\ensuremath{\mathbbm{1}}}
\newcommand{\ul}{\underline}
\def\bseq{\begin{subequations}}
\def\eseq{\end{subequations}}

\numberwithin{equation}{section} %%
\usepackage[]{graphicx}

\def\d {{\rm d}}

\def\calc         {{\cal C}}

\def\cale         {{\cal E}}
\def\calf         {{\cal F}}
\def\calg         {{\cal G}}

\def\cali         {{\cal I}}

\def\calk         {{\cal K}}
\def\call         {{\cal L}}
\def\calm         {{\cal M}}
\def\caln         {{\cal N}}

\def\calp         {{\cal P}}

\def\cals         {{\cal S}}

\def\calv         {{\cal V}}

\def\del          {\partial}

\def\ii           {{\rm i}}

\def\Re           {{\rm Re\hskip0.1em}}
\def\Im           {{\rm Im\hskip0.1em}}

\def\sqr#1#2{{\vcenter{\vbox{\hrule height.#2pt
 \hbox{\vrule width.#2pt height#1pt \kern#1pt \vrule width.#2pt}\hrule
 height.#2pt}}}}

\def\d{\text{d}}

\def\slashchar#1{\setbox0=\hbox{$#1$}           % set a box for #1
\dimen0=\wd0                                 % and get its size
\setbox1=\hbox{/} \dimen1=\wd1               % get siste of /
\ifdim\dimen0>\dimen1                        % #1 is bigger
\rlap{\hbox to \dimen0{\hfil/\hfil}}      % so center / in box
#1                                        % and print #1
\else                                        % / is bigger
\rlap{\hbox to \dimen1{\hfil$#1$\hfil}}   % so center #1
/                                         % and print /
\fi}

\begin{document}
\font\cmss=cmss10 \font\cmsss=cmss10 at 7pt

\vskip -0.5cm
%\rightline{\small{\tt MPP-2007-162}}
\rightline{\small{\tt ROM2F/2011/13}}

\vskip .7 cm
%\hfill IC/2004/ \vskip .1in \hfill CPHT \vskip .1in \hfill hep-th/yymmnnn

\hfill
\vspace{18pt}
\begin{center}
{\Large \textbf{Electrified branes}}
\end{center}

\vspace{6pt}
\begin{center}
{\large\textsl{ Luca Martucci}}

\vspace{25pt}
%\textit{\small $^a$ Max-Planck-Institut f\"{u}r Physik -- Theorie,\\
 %                   F\"{o}hringer Ring 6,  D-80805 M\"{u}nchen, Germany}\\ \vspace{6pt}
\textit{\small I.N.F.N. Sezione di Roma ``TorVergata'' \&\\  Dipartimento di Fisica, Universit\`a di Roma ``TorVergata", \\
Via della Ricerca ScientiÞca, 00133 Roma, Italy }\\  \vspace{6pt}
%\textit{\small Arnold Sommerfeld Center for Theoretical Physics,\\ LMU M\"unchen,
%Theresienstra\ss e 37, D-80333 M\"unchen, Germany}\\  \vspace{6pt}
\end{center}

\vspace{12pt}

\begin{center}
\textbf{Abstract}
\end{center}

\vspace{4pt} {\small \noindent A geometrical form of the supersymmetry conditions for D-branes 
on arbitrary type II supersymmetric backgrounds is derived, as well as the associated  BPS bounds.
The treatment is general and allows to consider, for instance,  non-static configurations or D-branes
supporting a non-vanishing electric flux, hence completing previous partial results.
In particular, our discussion  clarifies how the notion of calibration can be extended in order to be applicable to the most general
supersymmetric configurations. As  an exemplifying preliminary step,  the  procedure followed is first applied to  fundamental strings.

\noindent }

\vspace{1cm}

%\noindent {\em Possible comment ..........................................................................................................................................}

\thispagestyle{empty}

\vfill
\vskip 5.mm
\hrule width 5.cm
\vskip 2.mm
{\small
\noindent e-mail:  luca.martucci@roma2.infn.it, 
}

\newpage

\setcounter{footnote}{0}

\tableofcontents

%%%%%%%%%%%%%%%%%%%%%%%%%%%%%%%%%%%%%%%%

\vspace{1cm}
 %%%%%%%%%%%%%%%%%%%%%%%%%%%%%%%%%%%%%%%%%

\section{Introduction}

String theory is populated by branes of various kinds and their properties are directly related to the supersymmetric structure of the theory. 
This deep interplay has one of its most explicit manifestations in the appearance of integrable background geometrical structures  which are directly related to the  geometry of  branes. This relation constitutes an important ingredient in several applications. For instance, the presence of an integrable complex structure is naturally associated with holomorphically embedded branes. One can then often avoid facing the detailed supersymmetric structure, working directly with its physically relevant geometrical manifestation.

The relation between branes and background geometry  can be concretely realized in terms of calibrations, background forms with particular algebraic and differential properties -- see for instance \cite{gauntlettreview} for a review emphasizing some of the aspects that are important  for  applications to string theory. In their original mathematical definition \cite{HL}, calibrations are closed forms which identify  volume minimizing submanifolds with special properties.  They appeared in string theory when physicists started studying string theory vacua on Ricci-flat spaces with special holonomy (i.e.\ with reduced supersymmetry),  and the associated brane configurations -- see for instance  \cite{joyce}.  Physically, brane energy is expected to be subject to a BPS lower bound saturated  by supersymmetric configurations and, in string theory vacua on special holonomy spaces,  the energy of static branes is just given by the volume of the cycle wrapped by the brane. On the other hand, special holonomy spaces  are naturally equipped with calibrations and indeed these provide a nice  geometrical realization of the expected BPS energetic bound, see e.g. \cite{BBS}. 

Clearly, more complicated backgrounds require a modification of the notion of calibration given in \cite{HL}. 
Still focusing on static settings, the inclusion of background fluxes which couple minimally to  branes has been considered in \cite{GPT}
and leads to the general receipt  that the associated background calibrations are no-longer closed but rather
their exterior derivative is given by the background flux itself. After this work, several papers followed developing this idea at various levels of 
generality, see for instance \cite{marte,smith04,ura,saffin,papa06}.

On the other hand various branes, and in particular D-branes, can support non-trivial world-volume fluxes, which contribute to the brane energetics and need to be taken into account in a proper analysis. The way to incorporate magnetic world-volume fluxes for static D-branes on static backgrounds has been identified in \cite{paulcal,lucal,jarahluca,lucapaul} and exhibits the relevance of generalized geometry \cite{hitchin,gualtieri}\footnote{For a review for physicists, see for instance \cite{paulreview}.} as the proper framework for describing not only the D-brane geometry but also their dynamics \cite{lucasup,lucapauldef}. Not surprisingly, the results on the D-brane side uniform well with results the on the closed string sector side \cite{gmpt,glw,bg,tentofour,warpedeff,Haack:2009jg,luestdimi}. Importantly, it has also become evident that the calibration structures play a crucial role in  proving the integrability properties of supergravity backgrounds  fully coupled to backreacting D-branes and orientifolds, not only in supersymmetric settings \cite{dimipaul} but also in non-supersymmetric backgrounds which 
preserve some BPS-like properties \cite{dwsb}. 

This paper studies  the geometrical properties of general supersymmetric   D-branes on general supersymmetric type II 
backgrounds. In particular, backgrounds as well as D-branes can be non-static and the latter can support an electric flux in addition to the magnetic one.\footnote{The analogous problem for M5-branes has been addressed in \cite{lambert}.} The analysis explicitly confirms generalized geometry as the natural language to be used and indeed we will see
how the background supersymmetry conditions found in \cite{ale11} within this same framework 
are the ones which naturally enter in  the geometric characterization of  supersymmetric D-branes and the associated BPS bounds. 
In particular, our discussion shows how  various notions of calibrations should be extended in order to uncover non-static configurations 
and world-volume electric fluxes.  Furthermore, supersymmetry may be broken still preserving the calibrations and the associated BPS structures, along the lines of what was discussed for static settings in \cite{dwsb}. In this paper we will not consider this interesting possibility and the adjectives  ``BPS" and  ``supersymmetric" can be considered as synonymous.

This paper is structured as follows. After a brief summary in section \ref{sec:back} of the background properties, in section \ref{sec:F1} we start by considering the simpler case of fundamental strings. This will allow us to explain without too many technical complications the basic conceptual strategy which we will follow when we will turn to D-branes.  These are discussed in section \ref{sec:Dbranes} along the same lines followed in section \ref{sec:F1} for fundamental strings. In section \ref{sec:examples} we provide a few simple explicit  examples in which the we apply the general results of the previous sections.

%%%%%%%%%%%%%%%%%%%%%%%%%%%%%%%%%%%%%%%%%%%%%%%%%
%%%%%%%%%%%%%%%%%%%%%%%%%%%%%%%%%%%%%%%%%%%%%%%%% 

\section{Background structures}
\label{sec:back}

This section briefly  reviews the geometric structures characterizing the background. 

Take a supersymmetric type II vacuum, hence characterized by a pair of ten-dimensional Majorana-Weyl spinors ($\epsilon_1,\epsilon_2$)
which satisfy the supersymmetric Killing spinor equations.  In our conventions $\epsilon_1$ has positive chirality, $\Gamma_{11}\epsilon_1=\epsilon_1$, while $\Gamma_{11}\epsilon_2=\mp\epsilon_2$ in IIA/IIB respectively. These can be grouped into the Killing doubled-spinor
\be\label{ds}
\epsilon:=\left(\begin{array}{c}\epsilon_1 \\ \epsilon_2\end{array}\right)
\ee
The supersymmetric Killing spinor conditions have been analyzed in several papers, see for instance \cite{marte,smith04,saffin,ulf,ale11}. For our purposes, the most natural approach is the one adopted \cite{ale11}, which we follow.

By using $\epsilon$ one can construct several tensorial quantities, which have special differential properties deriving from the Killing spinor equations.  First of all, one can construct a vector field
\be\label{defK}
K:=-\frac12\bar\epsilon\,\Gamma^M\epsilon\,\partial_M\equiv-\frac12(\bar\epsilon_{1}\Gamma^M\epsilon_{1}+\bar\epsilon_{2}\Gamma^M\epsilon_{2})\,\partial_M
\ee
where we use capital letters from the middle of the alphabet as ten-dimensional curved indices.
By using the Killing spinor equations, $K$ turns out to be a Killing vector
\be\label{invmetric}
\call_Kg=0
\ee
and actually generates a symmetry of the entire background, hence including all bosonic supergravity fields and  supersymmetric structures. 
This will be important for us, since $K$  will identify the naturally selected `energy' which is minimized by the supersymmetric branes. In general 
$K$ can be either time-like or null and the overall minus-sign in (\ref{defK}) has been introduced in order for $K$ to be future-pointing.

In addition to the vector $K$, one can introduce a one-form $\omega$ defined by
\be\label{defomega}
\omega:=-\frac12\bar\epsilon\,\Gamma_M\sigma_3\epsilon\,\d x^M\equiv-\frac12(\bar\epsilon_{1}\Gamma_M\epsilon_{1}-\bar\epsilon_{2}\Gamma_M\epsilon_{2})\,\d x^M
\ee
Notice that $g(K, K)=-g^{-1}(\omega, \omega)\geq 0$, which follows from the fact that the vectors $\bar\epsilon_{1}\Gamma^M\epsilon_1$ and  $\bar\epsilon_{2}\Gamma^M\epsilon_2$ are null, so that either $K$ and $\omega$ are both null or $K$ is time-like and $\omega$ space-like. 
Furthermore, $\omega$ is related to $K$ by the differential condition
\be\label{Keq}
\d \omega=-\iota_KH
\ee
where $H$ is the (closed) NS-NS three-form which is locally given by the exterior derivative of the $B$-field, $H=\d B$.

There is one further set of tensors which can be constructed by using $\epsilon$. These can be organized into a 
 `polyform'
\be
\Psi=\sum_k\Psi_{\it (k)}
\ee
with ${\it k}=$even/odd in IIA/IIB and
\be
\Psi_{\it (k)}=-\frac{1}{{\it k}!}(\bar\epsilon_1\Gamma_{M_1\ldots M_k}\epsilon_2)\, \d x^{M_1}\wedge\ldots \d x^{M_k} 
\ee
Alternatively, $\Psi$ can be defined by the  bi-spinor  $\slashed\Psi=32\epsilon_1\otimes\bar\epsilon_2\Gamma_{11}$.\footnote{We use the supergravity and D-brane conventions described in appendix of the first paper of Ref.~\cite{dwsb} and in \cite{dirac}. Our conventions and definitions slightly differ from those of \cite{ale11}. The dictionary is the following: $H_{\rm here}=-H_{\rm there}$, $F^{\rm IIA}_{\rm here}=-F^{\rm IIA}_{\rm there}$,  $F^{\rm IIB}_{\rm here}=F^{\rm IIB}_{\rm there}$, $K_{\rm here}=-32K_{\rm there}$, $\omega_{\rm here}=-32\tilde K_{\rm there}$, $\Psi^{\rm IIB}_{\rm here}=-32\Phi^{\rm IIB}_{\rm there}$,  $\Psi^{\rm IIA}_{\rm here}=32\Phi^{\rm IIA}_{\rm there}$.}
Notice that $\Psi$ obeys the self-duality relation 
\be\label{selfpsi}
\Psi=*\lambda(\Psi)
\ee
where $\lambda(\Psi_{(k)})=(-)^{\frac{k(k-1)}{2}}\Psi_{(k)}$. Furthermore
\be\label{annV}
\iota_{K}\Psi+\omega\wedge \Psi=0
\ee

The Killing spinor equations imply that $\Psi$ is linked with the R-R fluxes by the following differential equations
\be\label{10PSeq}
\d_H( e^{-\phi}\,\Psi)=-\iota_{K}F-\omega\wedge F
\ee
where \be
F:=\sum_{\it k}F_{\it (k)}
\ee
is the sum of (even/odd) R-R field strengths (in IIA/IIB) and we have introduced the $H$-twisted exterior derivative 
\be
\d_H:=\d+H\wedge
\ee
which acts on polyforms. Notice that $\d_H^2=0$ by the Bianchi identity $\d H=0$.
The R-R field strengths are subject to the same self-duality constraint of $\Psi$: $F=*\lambda(F)$.

Not only the supergravity bosonic fields are invariant under the diffeomorphism generated by $K$, but also the supersymmetric spinorial structure. This implies in particular that
\be
\call_{K}\omega=0\, ,\quad \call_{K}\Psi=0
\ee
as can be derived from (\ref{Keq}) and (\ref{10PSeq}) by using the Bianchi identities.

Notice that the differential conditions reviewed above are necessary but not generically sufficient for guaranteeing the background  to be supersymmetry. Rather, they must  be supplemented by other conditions \cite{ale11}. However, the differential conditions above are the only ones that we will need when discussing fundamental strings and D-branes.

\bigskip

In discussing D-branes, it will be much more natural to use the language of generalized geometry \cite{hitchin,gualtieri}, see \cite{paulreview} for a review.
In particular,  the combination
\be\label{genK}
\calk=K+\omega
\ee
can be seen as a generalized vector in $TM\oplus T^*M$, known as generalized tangent bundle of the ten-dimensional manifold $M$. On this  there exists a natural inner product defined by
\be
\cali(X,X)=\iota_v\eta\qquad \text{for}\qquad  X= v+\eta\in TM\oplus T^*M
\ee
which induces a canonical $O(10,10)$ structure on $TM\oplus T^*M$. Polyforms can be seen as $O(10,10)$ spinors and the Clifford action of a generalized vector $X=v+\eta$ on them is given by the operator 
\be\label{cliffaction}
X\cdot := \iota_v +\eta\wedge
\ee

Furthermore, by using the ordinary metric $g$, one can introduce the generalized metric\footnote{We are using the SO(10,10) frame in which there are no off-diagonal elements in $\calg$. They can be obtained by a general $b$-twist which would however shift $H\rightarrow H-\d b$ -- see for instance \cite{paulreview}.}
\be\label{genG}
\calg=\left(\begin{array}{cc}
g & 0 \\
0 & g^{-1}
\end{array}\right)
\ee
which further reduces the generalized structure group to $SO(1,9)\times SO(1,9)$.

$\calk$ is null in both $\cali$ and $\calg$: 
\be\label{nullv}
\cali(\calk,\calk)=\calg(\calk,\calk)=0
\ee
Furthermore, $\Psi$ can be seen as a spinor in $TM\oplus T^*M$ and  (\ref{annV}) can be translated by saying that  $\Psi$ 
is annihilated by the  Clifford action of $\calk$:
\be
\calk\cdot \Psi=0
\ee
By adopting this viewpoint, one can write (\ref{10PSeq}) as
\be\label{gen10PS}
\d_H(e^{-\phi}\Psi)=-\calk\cdot F
\ee
and check that $\calk$ generates a generalized isometry which preserves the bosonic background as well as the spinorial structure. 
For instance
\bea
&&\call^H_\calk(e^{-\phi}\Psi)=0\\
&&\call^H_\calk F=\calk\cdot\d_HF=0 \label{locs}
\eea
where we have used the Bianchi identity $\d_H F=0$ -- we will later come back on the effect of localized sources -- and the generalized Lie derivative on forms is defined by
\be
\call^H_\calk:=\{\d_H,\calk\}
\ee

In conclusion,  $\calk$ can be considered as a generalized Killing vector.  $\calk$ and $\Psi$ will play an important role when discussing the energetics of D-branes. 

%%%%%%%%%%%%%%%%%%%%%%%%%%%%%%%%%%%%%%%%%%%%%%%%%%%%%%%%%
%%%%%%%%%%%%%%%%%%%%%%%%%%%%%%%%%%%%%%%%%%%%%%%%%%%%%%%

\section{Fundamental strings}
\label{sec:F1}

In this section we consider  fundamental (F1) strings in general supersymmetric
 backgrounds, by using the framework described in section \ref{sec:back}. We study the supersymmetry conditions and we derive the associated BPS bound and its relation to background calibrations. This section provides also a useful warm-up for the discussion on D-branes presented in section \ref{sec:Dbranes}, which will follow a path which is conceptually identical  although slightly more difficult from the technical point of view. However, this section and section \ref{sec:Dbranes} are almost completely independent and the reader more interested on D-branes may  safely jump directly to section \ref{sec:Dbranes}.

Let us briefly review some basic facts about the F1 world-volume theory, basically to fix our notation and conventions.
The action is given by
\be
S_{\rm F1}=-\mu_{\rm F1}\int_\cals\d^2\xi\sqrt{-\det g|_\cals}\,+\mu_{\rm F1}\int_\cals B
\ee
where $\xi^\alpha$ are world-sheet coordinates, $\mu_{\rm F1}=1/(2\pi\alpha')$, $\cals$ is the two-dimensional submanifold wrapped by the F1-string and $g|_\cals$ refers to the pull-back of bulk metric to $\cals$. In the following we will sometimes use an economic 
notation  for pulled-back tensors in components, leaving implicit the pull-back operation $|_\cals$.  For instance,  $B_{\alpha\beta}\equiv B_{MN}\partial_\alpha x^M \partial_\beta x^N\equiv (B|_\cals)_{\alpha\beta}$.

It will result useful to explicitly split  the action in terms of the Nambu-Goto and Wess-Zumino terms: $S=\int\d^2\xi\, \call=\int\d^2\xi(\call_{\rm NG}+\call_{\rm WZ})$, with
\be\label{F1lagr}
\call_{\rm NG}=-\mu_{\rm F1}\sqrt{-\det g|_\cals}\, ,\qquad \call_{\rm WZ}=\frac{\mu_{\rm F1}}{2}\,\epsilon^{\alpha\beta}B_{\alpha\beta}
\ee

%%%%%%%%%%%%%%%%%%%%%%%%%%%%%%%%%%%%%%%%%%%%%%%%%%%%%%%%%%%
%%%%%%%%%%%%%%%%%%%%%%%%%%%%%%%%%%%%%%%%%%%%%%%%%%%%%

\subsection{Geometric supersymmetry conditions}
\label{sec:F1cov}

We now study  the conditions that need to be imposed on supersymmetric fundamental strings and how they can be written in terms of the background tensors characterizing the bulk supersymmetry.

Let us start from the supersymmetry condition in spinorial form. This reads
\be\label{F1spinsusy}
\Gamma_{\rm F1}\epsilon=\epsilon
\ee
where $\epsilon$ is the doubled-spinor introduced in (\ref{ds}) and $\Gamma_{\rm F1}$ is the $\kappa$-symmetry operator \cite{superstring}
\be
\Gamma_{\rm F1}=\frac{1}{2\sqrt{-\det g|_\cals}}\,\epsilon^{\alpha\beta}\Gamma_{\alpha\beta}\otimes \sigma_3
\ee
where, according to our general conventions, $\Gamma_{\alpha\beta}=\Gamma_{MN}\partial_\alpha x^M \partial_\beta x^N$.
Our aim is to rewrite (\ref{F1spinsusy}) in a more geometrical way. This is readily obtained just by observing that (\ref{F1spinsusy}) implies that $\bar\epsilon\,\Gamma^M\Gamma_{\rm F1}\epsilon=\bar\epsilon\,\Gamma^M\epsilon$. In turn, by using the definitions (\ref{defK}) and (\ref{defomega}), it is easy to see that this latter condition can be written as 
\begin{equation}
\label{F1form2}
(\d X^M\wedge \omega)|_\cals=K^M\sqrt{-\det g|_\cals}\,\d^2\xi
\end{equation}

We then find that (\ref{F1form2}) is a {\em necessary} condition for the F1-string to be supersymmetric. On the other hand, we will show in the following sections that (\ref{F1form2}) is in fact also {\em sufficient} for it. Hence, (\ref{F1form2}) is completely equivalent
to (\ref{F1spinsusy}) and can be taken as defining supersymmetric F1-strings.\footnote{Actually, as we will see, (\ref{F1form2}) is redundant and only one particular `component' provides the minimal necessary and sufficient condition for the F1 to be BPS.}

We can write (\ref{F1form2}) in an alternative way. Consider the vector field $V=V^M\partial_M$ defined by
\be\label{defV}
V^M=\frac{\mu_{{\rm F}1}}{2\sqrt{-\det g|_\cals}}\,  \epsilon^{\alpha\beta}(\d X^M\wedge \omega)_{\alpha\beta}
\ee
$V$ is tangent to the F1 world-sheet, $V\in T\cals\subset (TM)|_\cals$. Hence the supersymmetry condition (\ref{F1form2}) states that $V$
must be equal to the Killing vector:
\be\label{F1susy}
V=K\qquad \text{(on $\cals$)}
\ee

Cleary, (\ref{F1susy})  implies that $K$ is tangent to the world-sheet $\cals$ of supersymmetric F1-strings. This means that $\cals$ does not break the symmetry generated by $K$, as expected. This condition can be alternatively formulated by saying that the embedding $\cals$ is a {\em stationary} configuration with respect to the `time' $t$ (which can actually be either time-like or light-like) generated by $K=\partial_t$.
 In particular, (\ref{F1form2}) implies that for supersymmetric strings
\be
(\d t\wedge \omega)|_\cals=\sqrt{-\det g|_\cals}\,\d^2\xi
\ee
Hence,  for supersymmetric F1-strings we can write
\be\label{BPSactionF1}
S_{\rm F1}=-\mu_{\rm F1}\int_\cals(\d t\wedge\omega-B)
\ee 

On general grounds, supersymmetric configurations are expected to minimize some BPS-like lower bound on some appropriately defined energy. This is indeed true, as discussed in the following sections.

%%%%%%%%%%%%%%%%%%%%%%%%%%%%%%%%%%%%%%%%%%%%%%%%%%%%%%

\subsection{Conserved current and charge}
\label{sec:F1currents}

We first recall some well known results about symmetries and conserved currents.

Noether's theorem states that one can construct a conserved current associated with any  symmetry  under which the action is invariant. More explicitly, take a generic Lagrangian  $\call(\phi^I,\del\phi^I)$ and suppose that it is invariant, up to a total derivative $\partial_\alpha\Lambda^\alpha$, under an infinitesimal deformation $\delta\phi^I$:
\be\label{totder}
\delta\call=\partial_\alpha\Lambda^\alpha
\ee
Notice that $\Lambda^\alpha$ is defined up to divergence-less additional terms. 

Then, one can construct the current
\be
j^\alpha=-\delta\phi^I\,\frac{\partial\call}{\del(\del_\alpha\phi^I)}+\Lambda^\alpha
\ee
which  is conserved,  $\partial_\alpha j^\alpha=0$  (on-shell).

Let us now apply Noether's theorem to our problem. Take the transformation
\be
\delta_K x^M=K^M
\ee
which corresponds to shifting the fundamental string in the direction of the Killing vector  $K$ defined in (\ref{defK}). 
Consider first $\call_{\rm NG}$, see (\ref{F1lagr}). Being $\call_Kg=0$, we obviously
have $\delta_K\call_{\rm NG}=0$. Hence, we can construct the following associated current 
\be
j^\alpha_{\rm NG}=\mu_{\rm F1}\sqrt{-\det g|_\cals}\, g^{\alpha\beta}K_\beta
\ee

Consider now $\call_{\rm WZ}$. First notice that, by  (\ref{Keq}), we have
\bea\label{defB}
\call_KB&=&\iota_KH+\d(\iota_KB)\cr
&=& \d(\iota_KB-\omega)
\eea
Hence, we can still apply Noether's procedure but now we have to be careful since $\call_{\rm CS}$ is only invariant up to a total derivative. Comparing with (\ref{totder}), we can take
\be\label{lambdaF1}
\Lambda^\alpha_{\rm WZ}=\mu_{\rm F1}\,\epsilon^{\alpha\beta}(\iota_KB-\omega)_\beta
\ee
Hence, the corresponding current is given by
\bea
j^\alpha_{\rm WZ}&=&-\mu_{\rm F1}\,\epsilon^{\alpha\beta}(\iota_KB)_\beta+\Delta^\alpha_{\rm WZ}\cr
&=&-\mu_{\rm F1}\,\epsilon^{\alpha\beta}\omega_\beta
\eea
Let us stress that, in deriving $j^\alpha_{\rm WZ}$, we have crucially used the background supersymmetry condition (\ref{Keq}).

We can then conclude that the conserved current associated with the world-volume symmetry induced by the bulk Killing vector $K$ is given by the sum of $j^\alpha_{\rm NG}$ and $j^\alpha_{\rm WZ}$:
\be
j_{\rm F1}^\alpha=j^\alpha_{\rm NG}+j^\alpha_{\rm WZ}=\mu_{\rm F1}\big(\sqrt{-\det g|_\cals}\, g^{\alpha\beta}K_\beta-\epsilon^{\alpha\beta}\omega_\beta\big)
\ee

Suppose now to split the world-sheet coordinates as $\xi^\alpha=(\tau,\sigma)$. More precisely, consider  world-sheet embeddings  which have  structure $\cals=\mathbb{R}\times \Sigma$, where $\mathbb{R}$ parametrizes a time-like world-sheet coordinate $\tau$ whereas $\Sigma$ is a space-like one-dimensional submanifold parametrized by the space-like coordinate $\sigma$. 
Hence, we can use the conserved current $j^\alpha_{\rm F1}$ to define the conserved charge
\bea\label{DeltaF1}
\Delta_{\rm F1}&=&\int_\Sigma\d\Sigma_\alpha\, j^\alpha_{\rm F1}\equiv\int_\Sigma\d\sigma\, j^\tau_{\rm F1}\cr
&=&-\int_\Sigma\big[\d\sigma \, g(P,K)+\mu_{\rm F1}\,\omega\big]
\eea
where we have introduced the vector
\be\label{F1P}
P^M=-\mu_{\rm F1}\sqrt{-\det g|_\cals}\, g^{\alpha\tau}\partial_\alpha x^M\qquad\Rightarrow\qquad j^\tau_{\rm NG}=-g(P,K)
\ee
$P^M$ is a vector tangent to $\cals$ and can be identified with the `kinematical' gauge-invariant momentum conjugated to the embedding field $x^M(\xi)$.

%%%%%%%%%%%%%%%%%%%%%%%%%%%%%%%%%%%%%%%%%%%%%%%%%%%%%%

\subsection{Supersymmetry and BPS bound}
\label{sec:F1bound}

Experience with supersymmetric theories leads us to expect that  the energy  obeys a BPS lower bound which is saturated exactly when the F1-string is supersymmetric. Standard arguments are usually applied to static situations and lead to the identification of various kinds of background calibrations, which naturally realize the BPS bound.  

On the other hand, in our analysis we are considering the most generic supersymmetric background and F1-string configuration.   
Hence, we need to extend the usual approach to calibrations. As we now show, such a generalization is actually very natural. 
Furthermore, our procedure straightforwardly generalizes to other branes, as we will explicitly see for the case of D-branes.

Let us start again from the supersymmetry condition (\ref{F1spinsusy}). By applying the split $\cals=\mathbb{R}\times\Sigma$ we can write 
\be\label{F1split}
\Gamma_{\rm F1}= \frac{1}{\mu_{\rm F1} \sqrt{g|_\Sigma}}\slashed{P}\gamma_{\rm F1}=
-\frac{1}{\mu_{\rm F1} \sqrt{g|_\Sigma}}\gamma_{\rm F1}\slashed{P}
\ee 
where $\slashed{P}=P^M\Gamma_M$, with $P^M$ defined in (\ref{F1P}), and 
\be
\gamma_{\rm F1}=\frac{1}{\mu_{\rm F1} \sqrt{g|_\Sigma}}\,\Gamma_\sigma\otimes\sigma_3
\ee
%As an immediate consequence, ${\slashed{P}}\Gamma_{\rm F1}=- \Gamma_{\rm F1}{\slashed{P}}=-\mu_{\rm F1}\Gamma_\sigma\otimes \sigma_3$.

Furthermore, notice that if the string is not collapsed then  $C\slashed{P}$ is a {\em positive definite} matrix, in the sense that
\be\label{psF1}
\chi^TC\slashed{P}\chi\equiv \bar\chi \slashed{P}\chi > 0
\ee
for any (non-vanishing) doubled Majorana spinor $\chi$.  Indeed, from the definition (\ref{F1P}) it is easy to see that $P$ is time-like, $g(P,P)=-\mu_{\rm F1}^2 g_{\sigma\sigma}$, whenever the string is not collapsed. In the real representation that we are using, the charge-conjugation matrix is simply given by $C=\Gamma^{\ul 0}$. Being $P$ future pointing by construction, we can locally choose a frame $e_{\ul M}$ such that $e_{\ul 0}\propto P$, so that $C\slashed{P}=P^{\ul 0}\bbone> 0$, which immediately leads to (\ref{psF1}). Clearly, such a conclusion cannot depend on the choice of frame and then is valid in general.

We can now apply the general inequality (\ref{psF1}) to the spinor 
\be\label{F1psi}
\chi=(\bbone-\Gamma_{\rm F1})\epsilon
\ee 
Using $\Gamma^T_{\rm F1}=-C^{-1}\Gamma_{\rm F1}C$, a few manipulations allow to rewrite (\ref{psF1}) with (\ref{F1psi}) as
\be
\bar\epsilon\slashed{P}\epsilon\,\geq\, \bar\epsilon\slashed{P}\Gamma_{\rm F1}\epsilon
\ee
In turn, by taking into account (\ref{F1split}) and the definitions (\ref{defK}) and (\ref{defomega}), this can be rewritten in more geometric way as
\be\label{locboundF1}
-g(P,K)\,\d\sigma\geq \mu_{\rm F1}\,\omega|_\Sigma
\ee
which should be read $-g(P,K)\geq \mu_{\rm F1}\omega_\sigma$. 
Recalling (\ref{DeltaF1}), we see that (\ref{locboundF1}) provides exactly the expected local bound
\be\label{F1bound}
\Delta_{\rm F1}\geq 0
\ee 

On the other hand, it is clear by construction that (\ref{F1bound}) is saturated {\em if and only if}  the F1-string satisfies the supersymmetry condition (\ref{F1spinsusy}). Hence, the supersymmetry condition 
can  be alternatively written as
\be\label{F1BPS}
\text{SUSY F1-string}\qquad\Leftrightarrow\qquad -g(P,K)\,\d\sigma=\mu_{\rm F1}\,\omega|_\Sigma
\ee
We then conclude that $\omega$ can be seen as the proper generalization of the more standard calibrations for static backgrounds and static world-sheet configurations. In particular, (\ref{F1BPS}) substitutes the usual calibration bound for minimally coupled static branes \cite{GPT}, where the the l.h.s.\ simply reduces to the volume density. 

Notice that in order to interpret the {\em algebraic} bound (\ref{F1bound}) in terms of a conserved charge, a key role is played by the construction of the currents $j^\alpha_{\rm NG}$ and $j^\alpha_{\rm WZ}$, which in turn uses  {\em differential} background conditions. In particular for $j^\alpha_{\rm WZ}$ we have crucially used the bulk supersymmetry condition (\ref{Keq}). Without such condition $S_{\rm WZ}$ would not be invariant and no current could be constructed. In a sense, consistency of the F1 world-sheet theory with supersymmetry  demands that the background fields $\omega$ and $K$ must be related by the differential condition (\ref{Keq}).

The discussion followed in this section makes explicit use of the time-plus-space split, but can be easily related to the covariant condition presented in (\ref{F1form2}). Indeed, (\ref{F1BPS}) can be obtained as the equation resulting from the contraction of (\ref{F1form2}) with $P$. In terms of the vector $V$ introduced in (\ref{defV}), the BPS condition can be written as $g(P,K)=g(P,V)$ and 
the BPS bound (\ref{F1bound}) can be rephrased as
\be\label{F1boundm}
|g(P,K)|\geq |g(P,V)|
\ee
We then see that, as already anticipated, the conditions encoded in (\ref{F1form2}) are not just necessary but also sufficient for characterizing BPS F1-strings. Actually, just the projection of $(\ref{F1form2})$ along $P$ is sufficient and the remaining components in the orthogonal directions 
follow from this.

\subsection{BPS bound and energetics}
\label{sec:F1ham}

In order to relate the above analysis to a more standard Hamiltonian treatment,  pick-up a field $B$ such that   $\d\iota_K B=-\iota_K H=0$ or, in other words, $\call_K B=0$. This is consistent with the fact that $\call_K H=0$ and $\iota_K H$ is exact, as required by the supersymmetry condition (\ref{Keq}).  In this case $\delta_K\call_{\rm F1}$ is exactly invariant  and we can apply Noether's procedure with $\Lambda^\alpha_{\rm CS}\equiv 0$.\footnote{This particular choice of $B$ is equivalent to correcting $\Lambda^\alpha_{\rm F1}$ given in (\ref{lambdaF1}) by a divergence-less term which make it vanish -- see comment after (\ref{totder}).} It is then easy to see that the conserved charge is simply given by $-g(\hat P,K)$, the component along $K$ of  the canonical momentum
\be
\hat P_M=\frac{\del \call_{\rm F1}}{\del(\del_\tau x^M)}=P_M+\mu_{\rm F1}(\iota_M B)|_\Sigma
\ee
which, compared to the gauge-invariant  momentum $P_M$, incorporates the effect of the WZ-term too.
Hence, we get the conserved energy
\be
E_{\rm F1}:=-\int_\Sigma\d\sigma\,g(\hat P,K)
\ee

The local bound (\ref{locboundF1}) can now be translated into a BPS bound on the energy $E_{\rm F1}$
\be\label{energyBPS}
E_{\rm F1}\geq E^{\rm BPS}_{\rm F1}
\ee
where
\be\label{BPSenergyF1}
E^{\rm BPS}_{\rm F1}:=\int_\Sigma(\omega-\iota_K B)
\ee

Notice that, thanks to the background supersymmetry condition (\ref{Keq}), $ E^{\rm BPS}_{\rm F1}$ is a topological quantity which does not change under continuous deformations of $\Sigma$. 
Indeed, suppose that $\Sigma$ can be deformed to $\Sigma'=\Sigma+\partial\calc$, for some chain $\calc$. Hence, by using (\ref{Keq}) and the fact that $\d\iota_KB=-\iota_KH$, it is easy to see that
\be
\int_{\Sigma'} (\omega-\iota_K B)=\int_{\calc} [\d\omega-\d(\iota_K B)]+\int_{\Sigma} (\omega-\iota_K B)=\int_{\Sigma} (\omega-\iota_K B)
\ee
In this sense, $E^{\rm BPS}_{\rm F1}$ can be seen as the central charge which extremizes the energy canonically associated with the Killing vector $K$.   In fact, by (\ref{Keq}) we may pick-up $B$ such that $\iota_K B=\omega$. With this choice, $E_{\rm F1}\equiv \Delta_{\rm F1}$ and $E^{\rm BPS}_{\rm F1}\equiv 0$.

By imposing a partial static gauge in which $\tau$ is identified with the (time- or light-like) coordinate $t$ generated by $K$, $-g(\hat P,K)$ coincides with the canonical Hamiltonian density. In this case $E_{\rm F1}$ would indeed  coincide with the canonical energy, along the lines of what happens for (generalized) calibrations \cite{GPT} for static backgrounds and branes. See also \cite{smith04,ura,saffin} for analogous conclusions obtained by starting from the local type II supersymmetry algebra.

%In fact, it can be useful to restrict the attention to more general (not necessarily BPS) stationary F1-strings.  Explicitly, this  exists a world-volume vector $k^\alpha$ such that $K^M=k^\alpha \partial_\alpha x^M$. In this case $g(K,P)=-\mu_{\rm F1}\sqrt{-h}k^\tau$. Hence, if we impose a partial static gauge $\tau=t$, from the relation $\delta^M_t=k^\alpha\partial_\alpha x^M$ we obtain that $k^\tau=1$. Then, we can write the local bound (\ref{locboundF1}) as
%\be
%\sqrt{-h}\, \d\sigma\geq \omega|_\cals
%\ee
%which is just a more general version of the standard calibration bound. The energy of stationary configurations
%\be\label{sta_actionF1}
%E^{\rm stationary}_{\rm F1}=\mu_{\rm F1}\int_{\Sigma}\d\sigma\sqrt{-h}-\mu_{\rm F1}\int_{\Sigma}\iota_KB
%\ee  is then clearly minimized on BPS configurations, on which it take the value $\calc_{\rm F1}$ defined in (\ref{defCF1}). 

%%%%%%%%%%%%%%%%%%%%%%%%%%%%%%%%%%%%%%%%%%%%%%%%%%%%%%%%%
%%%%%%%%%%%%%%%%%%%%%%%%%%%%%%%%%%%%%%%%%%%%%%%%%%%%%%%

\section{D-branes}
\label{sec:Dbranes}

We now turn to D-branes. The line we will follow is practically identical to one  for F1-strings, up to some technical complication
due to the more sophisticated D-brane world-volume theory. 

The effective theory governing the bosonic sector of a D$p$-brane wrapping a $(p+1)$-cycle $\cals$ is given by Dirac-Born-Infeld (DBI) and Chern-Simons (CS) terms
\bea\label{Daction}
S&=&\int_\cals \d^{p+1}\xi \,\call_{\rm DBI}+\int_\cals\d^{p+1}\xi \, \call_{\rm CS}\cr
&=&-\mu_{{\rm D}p}\int_\cals \d^{p+1}\xi e^{-\phi}\sqrt{-\det (g|_\cals+\calf)}+\mu_{{\rm D}p}\int_\cals C\wedge e^{\calf}
\eea
where $\mu_{{\rm D}p}=(2\pi)^p(\alpha')^{-\frac{p+1}2}$ and $\calf$ is the gauge-invariant world-volume field-strength, which satisfies the Bianchi identity $\d\calf=H|_\cals$.

As we will see, generalized geometry will naturally enter the 
description and will be useful to make more manifest the formal analogy  between D-branes and F1-strings.

%%%%%%%%%%%%%%%%%%%%%%%%%%%%%%%%
\subsection{Geometric  supersymmetry conditions}
\label{sec:susyD}

As in the discussion for fundamental strings, let us start from the supersymmetry conditions for a D$p$-brane in its spinorial form (see e.g. \cite{kappa,dirac}) 
\be\label{Dkappasusy}
\Gamma_{{\rm D}p}\epsilon=\epsilon
\ee
where $\Gamma_{{\rm D}p}$ acts on type II doubled-spinors as
\be
\Gamma_{{\rm D}p}=\left(\begin{array}{cc}  0  & \hat\Gamma_{{\rm D}p} \\
\hat \Gamma^{-1}_{{\rm D}p} & 0\end{array}\right)
\ee
with
\be
\hat\Gamma_{\text{D}p}=\frac{1}{\sqrt{-\det(g|_\cals+\calf)}}\sum_{2l+s=p+1}\frac{\epsilon^{\alpha_1\ldots\alpha_{2l}\beta_1\ldots\beta_s}}{l!s!2^l}\calf_{\alpha_1\alpha_2}\cdots\calf_{\alpha_{2l-1}\alpha_{2l}}\Gamma_{\beta_1\ldots\beta_s}\ .
\ee
and $\hat\Gamma^{-1}_{{\rm D}p}=(-)^pC^{-1}\hat\Gamma^T_{{\rm D}p}C$. As for fundamental strings, we would like now to re-express (\ref{Dkappasusy}) in a more geometrical way, which highlights its physical implications.

 These can be obtained by noticing that the spinorial supersymmetry condition (\ref{Dkappasusy}) implies in particular $\bar\epsilon(\Gamma^M\Gamma_{{\rm D}p})\epsilon=\bar\epsilon\,\Gamma^M\epsilon$ and $\bar\epsilon(\Gamma_M\sigma_3\Gamma_{{\rm D}p})\epsilon=\bar\epsilon(\Gamma_M\sigma_3)\epsilon$, which can be translated as follows in terms of the bulk structures $K$, $\omega$ and $\Psi$:
\bea\label{calcond2}
[(\d X^M\wedge \Psi)|_\cals\wedge e^\calf]_{\rm top}&=&K^M\sqrt{-\det (g|_\cals+\calf)}\,\d^{p+1}\xi\cr
[(\iota_M\Psi)|_\cals\wedge e^\calf]_{\rm top}&=&\omega_M\sqrt{-\det (g|_\cals+\calf)}\,\d^{p+1}\xi
\eea
We then obtain that (\ref{calcond2}) are {\em necessary} conditions for the D-brane to be supersymmetric. In fact, as it will be clear from the discussion of the following sections, (\ref{calcond2}) are also {\em sufficient} and then (\ref{calcond2}) provides a complete, actually redundant,  set of conditions which a supersymmetric D-brane must satisfy.

With the help of generalized geometry one can rewrite (\ref{calcond2}) in more inspiring way. First take the following generalized vector in $(TM\oplus T^*M)|_\cals$:
\be\label{calv}
\calv=\frac{\epsilon^{\alpha_0\ldots\alpha_p}}{\sqrt{-\det \calm}}\big\{[(\d x^M\wedge \Psi)|_\cals\wedge e^\calf]_{\alpha_0\ldots\alpha_p} \partial_M+[(\iota_M \Psi)|_\cals\wedge e^\calf]_{\alpha_0\ldots\alpha_p} \d x^M\big\}
\ee
where we have introduced the shorthand notation
\be
\calm:= g|_\cals+\calf
\ee

This is the D-brane counterpart of the vector $V$ introduced in (\ref{defV}) for F1-strings.
It is easy to see that $\calv$  is tangent to $(\cals,\calf)$ in the generalized sense, i.e.\ it takes values in the D-brane generalized tangent bundle $T_{(\cals,\calf)}$, which is defined as \cite{gualtieri}
\be\label{gentang}
T_{(\cals,\calf)}=\{v+\eta\in T\cals\oplus T^*M|_\cals\, :\, \eta|_\cals=\iota_v\calf\}
\ee

Then, it is easy to see that the supersymmetry condition (\ref{calcond2}) is equivalent to 
\be\label{ggcond}
\calv=\calk
\ee 
where $\calk$ is given in (\ref{genK}).  It is clear that (\ref{ggcond}) necessary requires that
\be\label{tv}
\calk\in T_{(\cals,\calf)} 
\ee
This condition can be interpreted as the condition that the D-brane is stationary with respect to $\calk$. More in detail, this incorporates the following two conditions. First, $K$ is tangent to $\cals$, $K\in T\cals$, and then $\cals$ is invariant under the symmetry generated by $K$. Second, $K$ identifies an naturally selected electric flux component $\cale=\iota_K\calf$, which is completely fixed in terms of background one-form $\omega$:
\be\label{stationaryE}
\cale=\omega|_\cals
\ee
In particular, this implies that there cannot be  supersymmetric electric field $\cale=\iota_K\calf\neq 0$ if the background supersymmetry has $\omega=0$.\footnote{As an aside remark, let us mention that, although so far we have treated  D-branes as probes, the same conditions must hold in the case of backreacting D-branes too. Notice in particular that the condition (\ref{tv}) for all localized RR sources in the background is equivalent to the requirement that $\calk\cdot j_{\rm loc}=0$, where $ j_{\rm loc}\simeq \delta(\cals)\wedge e^{-\calf}$ is the current associated with backreacting D-branes. The same current appears on the r.h.s\ of the R-R Bianchi identity $\d_H F=j_{\rm loc}$, implying (\ref{locs}) even in presence of supersymmetric back-reacting D-branes.} From the Bianchi identity $\d\calf=H|_\cals$ and the bulk equation (\ref{Keq}) it immediately follows that $\calf$ is stationary too: $\call_K\calf=0$. 

Finally, notice that by introducing the `time' $t$ (which, actually, can be light-like) defined by $K=\partial_t$, the first of (\ref{calcond2}) implies the following value for the D-brane action evaluated on the BPS configuration 
\be
S^{\rm BPS}_{{\rm D}p}=-\mu_{{\rm D}p}\int_\cals(\d t\wedge e^{-\phi}\Psi-C)\wedge e^{\calf}
\ee
In the next sections we will discuss the  bounds associated with the BPS condition.

%%%%%%%%%%%%%%%%%%%%%%%%%%%%%%%%%%%%%%%%%%%%%%%%%%%%%%%%%

\subsection{Conserved current and charge}
\label{sec:Denergy}

In order to discuss the the BPS bound associated with the supersymmetry conditions we must identify the proper conserved current and charge that enter the bound. Of course, these quantities are associated with the Killing vector $K$. 

The bulk symmetry generated by $K$ induces a symmetry of would-volume theory.  The action of such a symmetry on the embedding world-volume  fields $x^M(\xi)$ is given by 
\be\label{trK}
\delta_K x^M=K^M
\ee
Being $\call_K g=0$, the pulled-back metric $g|_\cals$ is clearly invariant under (\ref{trK}). On the other hand, $B$ is not in general invariant under the symmetry generated by $K$, but rather we have (\ref{defB}). Hence $\calf$, which can be locally written  as $\d A+B|_\cals$,  is generically not automatically invariant under (\ref{trK}). On the other hand, it should be clear from (\ref{defB}) that we can compensate (\ref{trK}) by a  transformation  of the  world-volume gauge-field
\be\label{trA}
\delta_K A=(\omega-\iota_KB)|_\cals
\ee   
so that $\delta_K\calf=0$. 

With this definition of $\delta_KA$,  the DBI lagrangian $\call_{\rm DBI}$ in (\ref{Daction}) is manifestly invariant under the symmetry generated by $K$: $\delta_{K}\call_{\rm DBI}=0$. By following Noether's procedure, recalled in section \ref{sec:F1currents}, we can then construct the following current
\be
j^\alpha_{\rm DBI}= \mu_{{\rm D}p}\, e^{-\phi}\sqrt{-\det \calm}\,\big(\calm^{(\beta\alpha)}K_\beta+\calm^{[\beta\alpha]}\omega_\beta\big)
\ee

Take now the CS Lagrangian $\call_{\rm CS}$  in (\ref{Daction}). By using (\ref{10PSeq}), it is possible to see that $\call_{\rm CS}$ tranforms into a total derivative $\delta_K\call_{\rm CS}=\partial_\alpha\Lambda^\alpha_{\rm CS}$  under (\ref{trK}) and (\ref{trA}), with
\be
\Lambda^\alpha_{\rm CS}=\frac{\mu_{{\rm D}p}}{p!}\,\epsilon^{\alpha\beta_{1}\ldots\beta_p}[(\calk\cdot C-e^{-\phi}\Psi)|_\cals \wedge e^{\calf} \big]_{\beta_1\ldots\beta_p}
\ee 
where $\calk$ is the generalized Killing vector defined in (\ref{genK}) and $\calk\cdot$ acts as in (\ref{cliffaction}).
Then, Noether's procedure gives the current
\be
j^\alpha_{\rm CS}=  -\frac{\mu_{{\rm D}p}}{p!}\,\epsilon^{\alpha\beta_{1}\ldots\beta_p}[(e^{-\phi}\Psi)|_\cals \wedge e^{\calf} \big]_{\beta_1\ldots\beta_p}
\ee
Notice that the bulk differential condition (\ref{10PSeq}) has been crucial in deriving this expression.

The total conserved current is then given by
\be
j^\alpha_{{\rm D}p}=j^\alpha_{\rm DBI}+j^\alpha_{\rm CS}
\ee
Consider now a time-plus-space split $\cals=\mathbb{R}\times \Sigma$, with adapted coordinate $(\tau,\sigma^a)$, $a=1,\ldots,p$,
where $\Sigma$ is the space-like  surface ({\it alias} space-volume) spanning $\cals$, and let us define the split
\be\label{Fdec}
\calf=\d\tau\wedge \cale+\calf_{\rm mg}\quad\Rightarrow\quad\calf_{\rm mg}:=\calf|_\Sigma
\ee
in terms of the electric field $\cale$ and the magnetic component $\calf_{\rm mg}$. 

We can then construct the conserved charge
\bea\label{Dcharge}
\Delta_{{\rm D}p}&=&\int_\Sigma\d\Sigma_\alpha\, j^\alpha_{{\rm D}p}\equiv\int_\Sigma\d^{p}\sigma\, j^\tau_{{\rm D}p}\cr
&=&-\int_\Sigma\d^p\sigma \, \calg(\calp,\calk)-\mu_{{\rm D}p}\int_\Sigma (e^{-\phi}\Psi)|_\Sigma\wedge e^{\calf_{\rm mg}}
\eea
Here we have used  the generalized metric defined in (\ref{genG}).
Furthermore, in the last line we have introduced the `generalized momentum', which is a generalized vector defined by 
\be
\calp:=P^M\partial_M+g_{MN}\partial_\alpha  x^N\Pi^\alpha\d x^M
\ee
where 
\bea\label{def_mom}
P^M&:=& -\mu_{{\rm D}p}\,e^{-\phi}\sqrt{-\det \calm}\,(\calm^{-1})^{( \alpha\tau)}\partial_\alpha X^M\cr
\Pi^\alpha&:=&-\mu_{{\rm D}p}\,e^{-\phi}\sqrt{-\det \calm}\,(\calm^{-1})^{[\alpha\tau]}
\eea
can be considered as the kinematical gauge-invariant momenta associated with $x^M$ and $A_\alpha$.  It is easy to check that $\calp$ belongs  to the generalized tangent space $T_{(\cals,\calf)}$ defined in (\ref{gentang}) and, moreover,  for non collapsed D-branes it is always `time-like' with respect to the generalized metric $\calg$:
\be\label{sH}
\calg(\calp,\calp)=P^MP_M+h_{\alpha\beta}\Pi^\alpha\Pi^\beta=-\mu^2_p\, e^{-2\phi}\det m
\ee 
where
\be
m:=g|_\Sigma+\calf_{\rm mg}
\ee
The identity (\ref{sH}) corresponds to the superhamiltonian constraint originating from the world-volume time-like diffeomorphism invariance.\footnote{The supermomentum constraint, which is associated with the space-like diffeomorphism invariance, can be equally elegantly written as $\calp\in T^\perp_{(\Sigma,\calf_{\rm mg})}$, where $T^\perp_{(\Sigma,\calf_{\rm mg})}$ is the orthogonal complement of the generalized tangent bundle $T_{(\Sigma,\calf_{\rm mg})}$ in $(TM\oplus T^*M)|_\cals$ with respect to the generalized metric $\calg$.}
Furthermore,
\be\label{nullP}
\cali(\calp,\calp)= P_\alpha  \Pi^\alpha=0
\ee
consistently the fact that $\calp$ belongs to $T_{(\cals,\calf)}$, which  is maximally isotropic with respect to $\cali$.

We see that by using the language of generalized geometry the conserved charge $\Delta_{{\rm D}p}$ has a structure very similar to the corresponding conserved charge for F1-strings found in (\ref{DeltaF1}). As explained in section \ref{sec:F1bound}, $\omega$ can be interpreted as a calibration for fundamental strings. Indeed,  a similar interpretation holds for  $e^{-\phi}\Psi$ too, as we will presently explain.

\subsection{Supersymmetry and BPS bound}
\label{sec:Dbound}

We start by considering  the following operator acting on bi-spinors
\be
\slashed{\calp}:= P^M\Gamma_M+\Pi^\alpha\Gamma_\alpha\sigma_3=\left(\begin{array}{cc}   P^M\Gamma_M+\Pi^\alpha\Gamma_\alpha  & 0 \\
0&  P^M\Gamma_M-\Pi^\alpha\Gamma_\alpha\end{array}\right) 
\ee
The notation has been chosen on purpose, in order to highlight the analogy with the analysis for F1-strings.
By using (\ref{sH}) and (\ref{nullP}), one can immediately conclude that the vector fields $P^M\pm\Pi^\alpha\partial_\alpha x^M$ are time-like and future-ponting which, by the same reasoning presented in section \ref{sec:F1bound}, implies that
\be\label{gbound}
\chi (C\slashed{\calp})\chi>0
\ee
for any non-vanishing doubled-spinor $\chi$. We can now apply this inequality to
\be\label{psiD}
\chi=(1-\Gamma_{\rm Dp})\epsilon
\ee
in order to obtain a bound which is saturated exactly by  supersymmetric D-branes, i.e.\ when (\ref{Dkappasusy}) is satisfied.

The resulting bound can be written in physically more  meaningful form by considering a time-plus-space split $\cals=\mathbb{R}\times \Sigma$, with adapted coordinates $\xi^\alpha=(\tau,\sigma^a)$, $a=1,\ldots,p$. Observe that 
\be\label{splitD}
\Gamma_{{\rm D}p}=\frac{1}{\mu_{{\rm D}p}\, e^{-\phi}\sqrt{\det m}}\, \slashed{\calp}\,\gamma_{{\rm D}p}=\frac{(-)^p}{\mu_{{\rm D}p}\,e^{-\phi}\sqrt{\det m}}\,\gamma_{{\rm D}p}\, \slashed{\calp}
\ee
where
\bea
\gamma_{\text{D}p}&=&\left(\begin{array}{cc}  0  & \hat\gamma_{{\rm D}p} \\
(-)^{p+1}\hat \gamma^{-1}_{{\rm D}p} & 0\end{array}\right)\cr
&\text{with}&\hat\gamma_{\text{D}p}=\frac{1}{\sqrt{\det m}}\sum_{2l+s=p}\frac{\epsilon^{a_1\ldots a_{2l}b_1\ldots b_s}}{l!s!2^l}\calf_{a_1 a_2}\cdots\calf_{a_{2l-1}a_{2l}}\Gamma_{b_1\ldots b_s}\ .
\eea
One can then rewrite the inequality (\ref{gbound}) with $\chi$ given in (\ref{psiD}) as
\be
\bar\epsilon\,\slashed{\calp}\epsilon\geq \bar\epsilon\,\slashed{\calp}\Gamma_{{\rm D}p}\epsilon
\ee
which,  by (\ref{splitD}), in turn translates into the following bound involving the world-volume fields and the background polyform $\Psi$
\be\label{locbound2}
-\calg(\calp,\calk)\geq \frac{\mu_{{\rm D}p}}{p!}\epsilon^{\tau a_1\ldots a_p}[(e^{-\phi}\Psi)|_\Sigma\wedge e^{\calf_{\rm mg}}]_{a_1\ldots a_p}
\ee
or, in more intrinsic notation,
\be\label{locbound3}
-\calg(\calp,\calk)\, \d^p\sigma\geq \mu_{{\rm D}p}[(e^{-\phi}\Psi)|_\Sigma\wedge e^{\calf_{\rm mg}}]_{(p)}
\ee
This bound is saturated {\em if and only if} the D-brane is supersymmetric
\be\label{Dcalcond}
\text{SUSY D-brane}\quad \Leftrightarrow\quad -\calg(\calp,\calk)\, \d^p\sigma= \mu_{{\rm D}p}[(e^{-\phi}\Psi)|_\Sigma\wedge e^{\calf_{\rm mg}}]_{(p)}
\ee

Recalling (\ref{Dcharge}) we see that (\ref{locbound2}) translates into the lower bound on the conserved charge $\Delta_{{\rm D}p}$ associated with the symmetry $K$:
\be\label{DBPSbound}
\Delta_{{\rm D}p}\geq 0
\ee
This is exactly what we expected from the discussion of section \ref{sec:Denergy}! D-branes preserving the supersymmetry $\epsilon$
are exactly those that minimize the conserved charge $\Delta_{{\rm D}p}$. This is in line with what we expect from our experience on static backgrounds. In particular $e^{-\phi}\Psi$ can be seen as an extension of the generalized calibrations of \cite{paulcal,lucal,jarahluca,lucapaul} to the most general supersymmetric setting.

Let us stress once again that the bulk differential condition (\ref{10PSeq}) has been crucial for deriving this current. As it happened for fundamental strings, a consistent  D-brane world-volume theory seems to  `know' about (at least part of) the bulk differential conditions.

Finally, notice that the supersymmetry condition (\ref{Dcalcond}) is obtained by projecting (\ref{calcond2}) along $\calp$ by using the metric $\calg$. This proves the statement anticipated in section \ref{sec:susyD} that the conditions (\ref{calcond2}) are not only necessary but actually also sufficient for a D-brane to be supersymmetric.
Moreover, the reformulation in terms of the generalized vector $\calv$ defined in (\ref{calv}) allows us to write the local bound (\ref{Dcalcond}) as 
\be\label{DBPSbound2}
|\calg(\calp,\calk)|\geq |\calg(\calp,\calv)|
\ee
whose saturation reproduces the supersymmetry condition in the form (\ref{ggcond}).

\subsection{BPS bound and energetics}

As done for fundamental strings, in order to give a Hamiltonian interpretation of the above results, let us pick-up potential forms $B$ and $C$ that preserve the symmetry generated by $K$, namely $\call_K B=0$ and $\call_K C=0$. In this case the complete Lagrangian $\call_{{\rm D}p}$ is exactly invariant under the symmetry generated by $K$ and the conserved charge density is  $-K^M\hat P_M$, with
\bea
\hat P_M:=\frac{\del \call_{{\rm D}p}}{\del(\del_\tau x^M)}&=&P_M-\mu_{{\rm D}p}e^{-\phi}\sqrt{-\det \calm}\,(\calm^{-1})^{[\alpha\tau]}B_{MN}\partial_\alpha X^N
\cr&& +\frac{\mu_{{\rm D}p}}{p!}\,\epsilon^{\tau \alpha_1\ldots \alpha_p}\big[\iota_M (C\wedge e^B)\wedge e^{\calf-B}\big]_{\alpha_1\ldots \alpha_p}
\eea
The conserved charge is then given by the energy
\be
E_{{\rm D}p}=-\int_\Sigma\d^p\sigma\, g(\hat P,K)
\ee
and the bound (\ref{DBPSbound}) translates into
\be
E_{{\rm D}p}\geq E^{\rm BPS}_{{\rm D}p}
\ee
with
\be\label{bED}
E^{\rm BPS}_{{\rm D}p}=\int_\Sigma\d^p\sigma\, \hat \Pi^a(\omega-\iota_K B)_a+\mu_{{\rm D}p}\,\int_\Sigma (e^{-\phi}\Psi-\calk\cdot C)\wedge e^{\calf_{\rm mg}} 
\ee
where we have introduced the momenta conjugated to the gauge field
\be\label{hatpi}
\hat \Pi^\alpha:=\frac{\del \call_{{\rm D}p}}{\del(\del_\tau A_\alpha)}=\Pi^\alpha +\frac{\mu_{{\rm D}p}}{(p-1)!}\,\epsilon^{\tau \alpha \beta_1\ldots \beta_{p-1}}\big[C\wedge e^{\calf_{\rm mg}}\big]_{\beta_1\ldots \beta_{p-1}}
\ee
These satisy the constraint $\hat \Pi^\tau\equiv 0$ and the Gauss law constraint $\del_\alpha\hat\Pi^\alpha\equiv \partial_a\hat\Pi^a=0$.

Remarkably, $E^{\rm BPS}_{{\rm D}p}$ is a topological quantity, which is invariant under continuous deformations of the D-brane configuration. Let us discuss this point in some detail.

First consider the first term on the r.h.s.\ of (\ref{bED}) and rewrite it as
\be
\mu_{\rm F1}\int_\Sigma (\omega-\iota_K B)|_\Sigma\wedge \rho_{\rm F1}
\ee
where we have introduced the $(p-1)$-form $ \rho_{\rm F1}$ defined by
\be\label{F1diss}
(\rho_{\rm F1})_{a_1\ldots a_{p-1}}=2\pi \alpha'\,\hat\Pi^b\,\epsilon_{ba_1\ldots a_{p-1}}
\ee 
As the notation suggests, $\rho_{\rm F1}$ corresponds to the  density of fundamental strings `dissolved' on the world-space $\Sigma$. By the Gauss law constraint  $\partial_a\hat\Pi^a=0$, $\rho_{\rm F1}$ is closed and defines a cohomology class on $\Sigma$  which, once integrally quantized, is Poicar\'e  dual to the space-like one-cycle $\Sigma_{\rm F1}\subset \Sigma$ wrapped by the dissolved fundamental strings. Hence, we allow for continuous deformations of $\hat\Pi^a$ which does not change the homology class of the dual strings, i.e.\ which correspond to deformations of $\rho_{\rm F1}$ by an exact piece.  By (\ref{Keq}) and $\call_K B=0$, the one-form $\omega-\iota_K B$ is closed  and then we can write (\ref{F1diss}) as 
\be\label{effF1}
\mu_{\rm F1}\int_{\Sigma_{\rm F1}}(\omega-\iota_K B)
\ee 
Comparing with  (\ref{BPSenergyF1}), we see that it clearly reproduces the contribution of a BPS F1-string wrapping $\Sigma_{\rm F1}$.
Repeating the arguments following (\ref{BPSenergyF1}), (\ref{effF1}) is clearly invariant under continuous deformations of $\Sigma_{\rm F1}$ (as well as of its hosting cycle $\Sigma$) within the same homology class. 
  
Consider now the second term on the r.h.s.\ of (\ref{bED}) and notice that by $\call_k C=0$ and $\d_HC=F$ one can easily check that $\d_H(\calk\cdot C)=-\calk\cdot F$ and then, by  (\ref{10PSeq}), we also have $\d_H(e^{-\phi}\Psi-\calk\cdot C)=0$. This immediately implies that by continuously deforming $(\Sigma,\calf_{\rm mg})$ the second term on the r.h.s.\ of (\ref{bED}) does not change. Indeed, consider 
a $(p+1)$-dimensional  chain $\tilde\Sigma$ supporting $\tilde\calf$ and interpolating between two configurations $(\Sigma,\calf_{\rm mg})$ and $(\Sigma',\calf'_{\rm mg})$, i.e.\  such that $\del\calc=\Sigma-\Sigma'$, $\tilde\calf|_\Sigma=\calf_{\rm mg}$ and  $\tilde\calf|_{\Sigma'}=\calf'_{\rm mg}$. Then
\bea
\int_\Sigma (e^{-\phi}\Psi-\calk\cdot C)\wedge e^{\calf_{\rm mg}}&=&\int_{\Sigma'} (e^{-\phi}\Psi-\calk\cdot C)\wedge e^{\calf'_{\rm mg}}+\int_{\calc}\d[ (e^{-\phi}\Psi-\calk\cdot C)\wedge e^{\tilde\calf}]\cr
&=&\int_{\Sigma'} (e^{-\phi}\Psi-\calk\cdot C)\wedge e^{\calf'_{\rm mg}}
\eea
where in the last step we have used the Bianchi identity $\d\tilde\calf=H|_\cals$ and the bulk equation $\d_H(e^{-\phi}\Psi-\calk\cdot C)=0$.
Hence, the r.h.s.\ of (\ref{bED}) depends just on the generalized homology class \cite{jarahluca} of $(\Sigma,\calf_{\rm mg})$ and is, in this sense, topological.

In conclusion, we find that the energy of D-branes is (globally) minimized by the supersymmetric configurations.

%%%%%%%%%%%%%%%%%%%%%%%%%%%%%%%%
\subsection{Gauss law, Bianchi identity and equations of motion}
\label{sec:gauss}

So far we have focused on the supersymmetry conditions for D-branes. On the other hand, general on-shell D-brane configurations must satisfy a certain set of equations of motion and Bianchi identities. Here we would like to discuss which equations are automatic consequences of the supersymmetry condition and which
ones must be independently added.  

The equations of motion associated with the embedding scalar fields are automatically satisfied thanks to the energetics  arguments implied by the BPS bound discussed above. On the other hand, the story is not that simple when we turn to  the equations associated with the world-volume field-stregth $\calf$. These are the Bianchi identity\footnote{The delta-like $k$-form $\delta^{(k)}(M)$  on a $n$-dimensional manifold  $M$ associated with a $(n-k)$-dimensional submanifold $N\subset M$ is defined by the identity $\int_M\alpha\wedge \delta^{(k)}(M)\equiv \int_N\alpha $  for any $(n-k)$ smooth differential form $\alpha$ on $M$.}   
\be
\d\calf=H|_\cals+\delta^{(3)}(\partial\tilde\cals) 
\ee
where $\partial\tilde\cals\subset \cals$  is the boundary of the world-volume  $\tilde\cals$  of D$(p-2)$-branes ending on $\cals$,  and the equations of motion 
\be\label{gaugeEoM}
\partial_\beta\left(\frac{\del \call_{{\rm D}p}}{\del \calf_{\alpha\beta}}\right)=-\frac{\mu_{\rm F1}}{p!}\,\epsilon^{\alpha\beta_1\ldots \beta_p}\,\delta^{(p)}(\partial\cals_{\rm F1})_{\beta_1\ldots \beta_p}
\ee
where now $\partial\cals_{\rm F1}\subset \cals$  is the world-line which is the boundary  of the world-sheet  $\cals_{\rm F1}$  of F1-strings ending on $\cals$. 

For completeness,  we have indicated the possible delta-like contributions from localized sources, neglected so far.
The localized sources contributing to the the Bianchi identity are given by D$(p-2)$-branes ending on a codimension-three submanifold of $\cals$. The localized sources contributing to (\ref{gaugeEoM}) are given by the end-points of fundamental strings attached to the D-brane. 
Notice that the overall sign of the localized contributions depends on the orientation of open branes and strings ending on the hosting D-brane.

Let us now use the natural time-plus-space split introduced by the Killing vector $K=\partial_t$, by choosing the partial static gauge $\tau=t$. The split (\ref{Fdec}) in this partial static gauge reads
\be
\calf=\d t\wedge \cale+\calf_{\rm mg}
\ee
Analogously, we can split $H=\d t\wedge \iota_K H+H_{\rm mg}$.  Remember now the supersymmetry condition (\ref{stationaryE}).
By using (\ref{Keq}), it directly implies that the Bianchi identity for the electric field $\cale$ is {\em automatically} implied by the supersymmetry condition. Hence, it cannot receive any contribution from localized sources, which is consistent with the fact that the possible D$(p-2)$-branes wrapping  $\tilde\cals$ must be supersymmetric and then tangent to $K$ as well, hence giving  a vanishing localized contribution $\iota_K\delta^{(3)}(\partial\tilde\cals)\equiv 0 $. 

On the other hand, it  remains to independently  impose by hand the Bianchi identity for $\calf_{\rm mg}$. By splitting $\tilde\cals=\mathbb{R}\times \tilde\Sigma$, with $\del\tilde\Sigma\subset \Sigma$, we have
\be\label{sourceBI}
\d\calf_{\rm mg}=H_{\rm mg}|_\Sigma+\delta^{(3)}(\del\tilde\Sigma) 
\ee
 Consistency requires that the r.h.s.\ of (\ref{sourceBI}) is exact. Hence, if $H_{\rm mg}|_\Sigma$ is cohomologically non-trivial then $\Sigma$
must host the boundary of an appropriate number of D$(p-2)$-branes.

The situation is reversed for the equations of motion (\ref{gaugeEoM}). By splitting $\cals_{\rm F1}=\mathbb{R}\times \Sigma_{\rm F1}$, the time component of (\ref{gaugeEoM}) gives rise to the Gauss law 
constraint
\be\label{locgauss}
\partial_a\hat\Pi^a=-\frac{\mu_{\rm F1}}{p!}\,\epsilon^{b_1\ldots b_p}\,\delta^{(p)}(\partial\Sigma_{\rm F1})_{b_1\ldots b_p}
\ee
which must be independently imposed, in addition to the BPS conditions. It is easier to  describe the localized source terms entering (\ref{locgauss}) by using the dual description of $\hat\Pi^a$ provided by the $(p-1)$-form $\rho_{\rm F1}$ defined in (\ref{F1diss}).  $\rho_{\rm F1}$ is closed up to localized contributions coming from the end-points of F1-strings
\be\label{sourcegauss}
\d\rho_{\rm F1}=-\delta^{(p)}(\partial\Sigma_{\rm F1})
\ee

Notice that, by (\ref{hatpi}), $\rho_{\rm F1}$ receives a contribution from the R-R potentials, which could be not well defined globally. However, we can split $\rho_{\rm F1}=\tilde\rho_{\rm F1}-\frac{1}{(2\pi \ell_s)^{p}}\,[C|_\Sigma\wedge e^{\calf_{\rm mg}}]_{(p-1)}$, with $\ell_s:=\sqrt{\alpha'}$, where now $\tilde\rho_{\rm F1}$ is globally defined and is given by the same formula (\ref{F1diss}) with  $\Pi^a$ instead of $\hat\Pi^a$. Hence,  (\ref{sourcegauss}) can be written as
 \be
 \d\tilde\rho_{\rm F1}=\frac{1}{(2\pi \ell_s)^p}\,[F|_\Sigma\wedge e^{\calf_{\rm mg}}]_{(p)}-\delta^{(p)}(\partial\Sigma_{\rm F1})
 \ee 
Being  $\tilde\rho_{\rm F1}$ globally defined on $\Sigma$, consistency requires the r.h.s.\ of this equation to be exact. For instance, this is the reason why a D0-brane in a massive IIA background with $F_{(0)}=k$ requires $k$ attached F1-strings.

The fact that  Bianchi identity and Gauss law must be imposed by hand in addition to  the supersymmetry conditions should be also clear from the argument leading to the BPS energy bound, which uses them as external ingredients. On the other hand, the BPS energy bound itself guarantees that the remaining equations are automatically fulfilled.

Finally notice that, by using the source-modified equations (\ref{sourceBI}) and (\ref{sourcegauss}), one can extend the description of 
D-brane networks given in \cite{jarahluca} to our more general framework while incorporating fundamental strings too.  

%%%%%%%%%%%%%%%%%%%%%%%%%%%%%%%%
%%%%%%%%%%%%%%%%%%%%%%%%

%%%%%%%%%%%%%%%%%%%%%%%%%%%%%%%%%%%%%%%%%%%%%%%%%%%%
%%%%%%%%%%%%%%%%%%%%%%%%%%%%%%%%%%%%%%%%%%%%%%%%%%%%%%

\section{Some examples}
\label{sec:examples}

We now discuss a few examples of applications of the above general results.

\subsection{Magnetized D-branes}
\label{sec:mag}

In this section we consider  examples of D-branes with at most pure magnetic flux $\calf_{\rm mg}$ on static backgrounds.
 In the following two subsections, we focus on two subcases.

In the first subcase,  discussed in section \ref{sec:static}, $K$ is {\em timelike} and we identify it with the generator of the time characterizing the static background $K\equiv \partial_t$. In this case all supersymmetric branes, being tangent to $K$, must be static themselves and  the energetics governing these configurations can be characterized in terms of the generalized calibrations described in \cite{paulcal,lucal,jarahluca,lucapaul}. These are a direct extension  of the calibrations of \cite{HL} and \cite{GPT} which allows to incorporate the effect of  $\calf_{\rm mg}$. We  then show how to recover these cases from our general analysis.

In the second subcase  $K$ is {\em null} and then, in principle, supersymmetric D-branes can be either static (filling the 1+1 directions spanned by $\del_t$ and $K$) or travel at the speed of light along $K$.  In particular, $K$ is null in the interesting case of $\caln=1$ (static) compactifications to four dimensions and,  in section \ref{sec:compact},   we focus for concreteness on this important class of backgrounds.

\subsubsection{Static magnetized D-branes}
\label{sec:static}

Here we consider the case in which $K$ is time-like and the space is static. Hence, introducing the time $t$ defined by $K=\del_t$,  we can split the ten-dimensional space as $\mathbb{R}\times M_9$, with metric
\be
\d s^2_{(10)}=-e^{2A(y)}\d t^2+g_{mn}(y)\d y^m\d y^n
\ee 
Furthermore, since we want to focus on magnetized branes, we impose the condition $\omega=0$, which  implies that $H$ has only internal legs, $\iota_t H=0$. We can then split 
\be
\Psi=\psi+\d t\wedge \tilde\psi
\ee
where $\iota_t\psi=0$ and $\d t\wedge \tilde\psi=*\lambda(\psi)$ by (\ref{selfpsi}). The supersymmetry condition (\ref{10PSeq}) splits into 
\be
\d_H\psi=-\iota_t F \, ,\qquad \d_H\tilde\psi=0
\ee

Consider now a D$p$-brane wrapping $\cals=\mathbb{R}\times \Sigma$, with $\Sigma\subset M_9$, possibly supporting a non-trivial magnetic flux $\calf_{\rm mg}$, such that $\d\calf_{\rm mg}=H|_\Sigma$. By defining $m=g|_\Sigma+\calf_{\rm mg}$, we have $\sqrt{-\det \calm}=e^A\sqrt{\det m}$ and 
\be
P^M=\mu_{{\rm D}p}\, e^{-A-\phi}\sqrt{\det m} \, \delta^M_t\, ,\qquad \Pi^\alpha\equiv 0
\ee
Hence, $-\calg(P,\calk)=\mu_{{\rm D}p}\, e^{A-\phi}\sqrt{\det m}$ and the BPS bound $\Delta_{{\rm D}p}\geq 0$ given by (\ref{locbound3}) becomes
\be\label{staticbound}
e^A\sqrt{\det m}\,\d^p\sigma\,\geq\, [\psi|_\Sigma\wedge e^{\calf_{\rm mg}}]_{(p)}
\ee
This is saturated, i.e.\ the D$p$-brane is supersymmetric, exactly when $(\Sigma,\calf_{\rm mg})$ is calibrated by $\psi$.
This reproduces the result of the discussion for static configurations presented in \cite{lucapaul}. 

Of course, in the limit in which one turns off all background and world-volume fluxes, $e^A\sqrt{\det m}\,\d^p\sigma$ is just the volume form, $\psi$ reduces to a calibration of \cite{HL} and (\ref{staticbound}) gives the ordinary calibration bound on the volume of the cycle $\Sigma$.

\subsubsection{Compactifications to ${\rm Mink}_4$ and non-static branes}
\label{sec:compact}

A particularly important  class of static backgrounds is provided by the   $\caln=1$   compactifications to four-dimensional Minkowski space  with SU(3)$\times$SU(3) structure  considered in \cite{gmpt}, whose supersymmetry conditions can be completely interpreted in terms of (static) calibrations \cite{lucal}. Concerning the background structures, the relation between the general ten-dimensional approach and this particular sub-case is discussed in detail in \cite{ale11}.  

In this case the space has structure $\mathbb{R}^{1,3}\times M_6$, the general metric has the form
\be
\d s^2=e^{2A(y)}\d x^\mu\d x_\mu+\hat g_{mn}(y)\d y^m\d y^n
\ee
and all fields preserve the four-dimensional Poincar\'e invariance. We can split the ten-dimensional gamma-matrices as follows
\be
\Gamma_\mu=e^A\gamma_\mu\otimes \bbone\, ,\qquad \Gamma_m=\gamma_5\otimes \hat\gamma_m
\ee
where $\gamma_\mu$ are the four-dimensional gamma matrices for Minkowski space. 
By using the real representation in which $C=\Gamma^{\ul 0}$, the four independent Killing spinors (\ref{ds}) have the form
\be
\epsilon_1=\zeta\otimes \eta_1+\text{c.c.}\, ,\quad \epsilon_2=\zeta\otimes \eta_2+\text{c.c.}
\ee
where $\zeta$ is an arbitrary four-dimensional spinor of positive chirality $\gamma_5\zeta=\zeta$, while the internal spinors have chiralities  $\hat\gamma_7\eta_1=\eta_1$ and $\hat\gamma_7\eta_2=\mp\eta_2$ in IIA/IIB. We denote the norms of the internal spinors  by $|a|^2=\eta^\dagger_1\eta_1$ and $|b|^2=\eta^\dagger_2\eta_2$.

One can readily compute that $K^m=\omega_m=0$ while
\be
K^\mu=-e^{-A}\,\bar\zeta\gamma^\mu\zeta\, (|a|^2+|b|^2)\, ,\qquad \omega_\mu=-e^{A}\,\bar\zeta\gamma_\mu\zeta\, (|a|^2-|b|^2)
\ee
Furthermore, (\ref{invmetric}) implies that $\partial_m[e^{-A}(|a|^2+|b|^2)]=0$, while (\ref{Keq}) implies that $\partial_m[e^{A}(|a|^2-|b|^2)]=0$ since by Poincar\'e invariance $H$ has only internal legs.  

By (\ref{stationaryE}) the background is compatible with an electric flux (along $K$) only if $|a|^2\neq |b|^2$. On the other hand, in ordinary flux compactifications O-planes are usually necessary and they preserve the same supersymmetry  of non-fluxed D-branes, hence requiring 
$|a|^2=|b|^2$. Focusing on this latter case, we have $\omega=0$ and  we can choose $|a|^2=|b|^2=e^A$, so that
\be
K^\mu=-2\, \bar\zeta\gamma^\mu\zeta
\ee

We can use the real representation of the four-dimensional gamma matrices in which
\be
\gamma_0=\ii \sigma_2\otimes \bbone_2\, ,\quad \gamma_1=\sigma_1\otimes \bbone_2 \, ,\quad \gamma_2=\sigma_3\otimes \sigma_3\, ,\quad \gamma_3=\sigma_3\otimes \sigma_1
\ee
and the four-dimensional chirality operator is $\gamma_5:=\ii\gamma^{0123}=\bbone\otimes \sigma_2$, and pick up a particular $\zeta$ 
\be\label{speczeta}
\zeta=\frac{1}{2}\left(\begin{array}{c} 1 \\ 0\end{array}\right)\otimes \left(\begin{array}{c} 1 \\ \ii\end{array}\right)
\ee
We can then easily compute 
\be
K=\frac{\del}{\del t}+\frac{\del}{\del x^1}
\ee
We have used the notation $(t,x^i)\equiv x^\mu$ in order to emphasize the distinguished role of time. 
In turn, the polyform $\Psi$ becomes
\bea\label{compcal}
\Psi&=&(\d x^1-\d t)\wedge\, e^{2A}\Im\Phi_1+(\d x^1-\d t)\wedge \d x^2\wedge \d x^3\wedge\, e^{4A}\Re\Phi_1\cr
&&+(\d x^1-\d t)\wedge\, e^{3A}\Im\big[(\d x^2+\ii \d x^3)\wedge \Phi_2\big]
\eea
where we have introduced the internal pure spinors defined, through the Clifford map, by
\be\label{6ps}
\Phi_1:=-8\ii e^{-A}\, \eta_1\otimes \eta_2^\dagger\, ,\qquad \Phi_2:=\mp\ 8\ii  e^{-A}\, \eta_1\otimes \eta_2^T
\ee
the  upper/lower sign in $\Phi_2$ is for IIB/IIA respectively.

On the right-hand side of (\ref{compcal}) one can recognize the calibrations $\Re\Phi_1$, $\Im\Phi_1$ and $\Re(e^{\ii\theta}\Phi_2)$  (for some constant $e^{\ii \theta}$) on $M_6$ which were introduced in \cite{lucal} to describe {\rm static} space-filling, string-like and domain-wall D-branes respectively. 

For instance, a static   D$p$-brane which appears as a string in $\mathbb{R}^{1,3}$ and preserves (\ref{speczeta}) must be stretched along $x^1$ and wrap an internal magnetized $(p-1)$-cycle $(\hat\Sigma,\calf_{\rm mg})$ in $M_6$ calibrated by $\Im\Phi_1$:
\be\label{Dstringsusy}
[\Im\Phi_1|_{\hat\Sigma} \wedge e^{\calf_{\rm mg}}]_{(p-1)}=\sqrt{\det m}\, \d^{p-1}\sigma
\ee
where $m:= \hat g|_{\hat\Sigma}+\calf_{\rm mg}$.

More generically, if we want to consider more complicated static D-branes or D-brane networks, analogously to the case of time-like $K$ considered in section \ref{sec:static} we could actually write
\be
\Psi=\psi+*\lambda(\psi)
\ee
where
\bea
\psi&=&\d x^1\wedge\, e^{2A}\Im\Phi_1+\d x^1\wedge \d x^2\wedge \d x^3\wedge\, e^{4A}\Re\Phi_1\cr
&& +\d x^1\wedge\, e^{3A}\Im\big[(\d x^2+\ii \d x^3)\wedge \Phi_2\big]
\eea
is the calibration for static (possibly magnetized) branes on $\mathbb{R}^3\times M_6$ which was already considered in \cite{jarahluca,lucapaul}.\footnote{See in particular \cite{jarahluca} for a discussion on how D-brane networks can be described in this framework.} 

On the other hand, differently from the assumptions of section \ref{sec:static}, here $K$ is null and this opens the possibility of having 
non-static branes. A simple example of non-static supersymmetric D-branes is obtained as follows.
Consider a D-brane which appears as a (two-dimensional) string in $\mathbb{R}^{1,3}$ and wraps a internal magnetized cycle $(\hat\Sigma,\calf_{\rm mg})$ calibrated by $\Im\Phi_1$ as in
(\ref{Dstringsusy}). If $(\tau,\xi)$ are the two world-sheet coordinates along the string  in $\mathbb{R}^{1,3}$, take  an embedding
of the form
\be
t=\tau\, ,\quad x^1=\tau+f^1(\xi)\, ,\quad x^2=f^2(\xi)\, ,\quad x^3=f^3(\xi)
\ee
where $f^{1,2,3}$ are arbitrary functions. Such an embedding describes a string of {\em arbitrary} shape which travels at the speed of light in the $x^1$ direction. One can readily verify that $\sqrt{-\det \calm}=|\partial_\xi f^1|\sqrt{\det m}$ and then, by using the formula (\ref{calcond2}), one can easily check that all these configurations are supersymmetric.
Alternatively, one can compute $\calp$ from (\ref{def_mom}) and  obtain that
\be
-\calg(\calp,\calk)=-g(P,K)=\mu_{{\rm D}p}\, e^{-\phi}\,|\partial_\xi f^1|\sqrt{\det m}
\ee
On the other hand, we have that the $p$-dimensional world-space is given by $\Sigma=\gamma\times\hat\Sigma$, where $\gamma$ is the curve in $\mathbb{R}^3$ described by $x^i=f^i(\xi)$. Hence
\be
[\Psi|_\Sigma\wedge e^{\calf_{\rm mg}}]_{(p)}=(\partial_\xi f^1)\d \xi\wedge [\Im\Phi_1|_{\hat\Sigma} \wedge e^{\calf_{\rm mg}}]_{(p-1)} 
\ee
and by (\ref{Dstringsusy}) the supersymmetry condition in the form (\ref{Dcalcond}) is satisfied by appropriately choosing the orientation of $\xi$.

Similar supersymmetric configurations are known in the literature and the peculiar mechanism which allows them to be supersymmetric
is somewhat analogous to the one behind the existence of giant gravitons \cite{giant}, as discussed in \cite{mikhailov}.
Indeed, giant gravitons can be equally well described in our general formalism, although we refrain from explicitly discussing them in the present paper.\footnote{A proposal of calibration for giant gravitons appears in \cite{smith04}, which is consistent with the results of our approach.}

%%%%%%%%%%%%%%%%%%%%%%%%%%%%%%%%%%%%%%%%%%%%%%%%%%%%
%%%%%%%%%%%%%%%%%%%%%%%%%%%%%%%%%%%%%%%%%%%%%%%%%%%%%%

\subsection{Simple D-branes with non-vanishing electric field}
\label{sec:electr}

In the literature, there are several examples of BPS  D-branes supporting an electric field. 
In this section we revisit  a few simple known examples by exploiting the perspective offered by our general approach.
We will consider the BIon solution \cite{malda97} and the supertube solutions of \cite{mateos1,mateos2,mateos3,toni}.
Our discussion will be purposely slightly more general than in the original works, in order to highlight the power of our framework.

These examples can be seen as describing threshold bound states of D-branes and fundamental strings. Hence, this class of examples can be obtained by selcting backgrounds whose supersymmetry is compatible with supersymmetric F1-strings. A simple, still quite general class of backgrounds of this kind have the form $\mathbb{R}^{1,1}\times M_8$ with  $M_8$ any eight-dimensional (Ricci-flat) special holonomy space. We can write the ten-dimensional  metric as
\be
\d s^2_{M_{10}}=-\d t^2+\d x^2 +\hat g_{mn}(y)\d y^m\d y^n
\ee

These backgrounds preserve at least two supersymmetries. We are interested in the particular supersymmetry which is mutually supersymmetric with a fundamental string stretching along $\mathbb{R}^{1,1}$. We can describe it quite explicitly, by using 
a two-plus-eight split $x^M=(t,x,y^m)$ and accordingly decomposing the gamma matrices (in the real representation) as follows
\be
\Gamma_t=\ii\sigma_2\otimes \bbone\, ,\quad \Gamma_x=\sigma_1\otimes\bbone\,,\quad \Gamma_m=\sigma_3\otimes\hat\gamma_m
\ee
and $\hat\gamma_{(8)}=\hat\gamma_{1\ldots 8}$ as eight-dimensional chiral operator. 
Then, it is easy to see that the preserved supersymmetry has the form
\be
\epsilon_1=\left(\begin{array}{c} 1 \\ 0\end{array}\right)\otimes \eta_1\, \quad \epsilon_2=\left(\begin{array}{c} 0 \\ 1\end{array}\right)\otimes \eta_2
\ee
for $\eta_1$ and $\eta_2$ two Killing spinors on $M_8$, with $\gamma_{(8)}\eta_1=\eta_1$ and  $\gamma_{(8)}\eta_2=\pm\eta_2$ in IIA/IIB respectively. 
We can then easily compute the tensors characterizing this supersymmetry:
\be\label{eletrback}
K=\frac{\partial}{\del t}\, ,\quad \omega=\d x\, ,\quad \Psi=(1-\d t\wedge \d x)\wedge \psi 
\ee
where $\psi$ is the calibration for static magnetized D-branes (in the sense of \cite{paulcal,lucal,jarahluca,lucapaul} recalled in section \ref{sec:mag}) for D-branes filling the time direction $t$ and wrapping a cycle along $M_8$.  Take for instance IIA and $M_8$ to be a Spin(7) space, with Majorana-Weyl Killing spinor $\eta$, $\eta^T\eta=1$. Hence we can take $\eta_1=\eta_2=\eta$ and   $\psi=1+\Omega_{4}+\rm{vol}_8$, where $\Omega_{4}$ is the Cayley four-form, which is the calibration for the so-called Cayley cycles.

\subsubsection{BIons} 
\label{sec:bion}

In the space $\mathbb{R}^{1,1}\times M_8$ described above, consider a D$p$-brane with the following profile
\be
t=\tau\, ,\quad x=X(\sigma) \, ,\quad y^m=Y^m(\sigma)
\ee
More precisely, we take $\sigma^a$ to parametrize an internal cycle $\Sigma\subset M_8$, defined by the embedding $Y^m(\sigma)$, and
allow for a possible displacement of the embedding in the flat $x$ direction. 

From our general result (\ref{stationaryE}) we can immediately conclude that the electric field is fixed to be
\be
\cale=\del_aX\,\d\sigma^a
\ee
We then have
\be
\calm_{\alpha\beta}:=(g|_\cals+\calf)_{\alpha\beta}=\left(\begin{array}{cc} -1 & \nabla_b X \\ -\nabla_a X & \hat m_{ab}+\nabla_aX\nabla_bX\end{array}\right)
\ee
where $\hat m=\hat g|_\Sigma+\calf_{\rm mg}$. One then easily gets  $\det \calm=-\det \hat m$. 
On the other hand, by using (\ref{eletrback}) it is also easy to see that
\bea
&&[(\d t\wedge \Psi)|_\cals\wedge  e^\calf]_{(p+1)}=[(\iota_x\Psi)|_\cals\wedge  e^\calf]_{(p+1)}=\d t\wedge [\psi|_\Sigma \wedge  e^{\calf_{\rm mg}}]_{(p)}\cr
&&[(\d x\wedge \Psi)|_\cals\wedge  e^\calf]_{\rm top}=
[(\d y^m\wedge \Psi)|_\cals\wedge  e^\calf]_{\rm top}\cr &&
=[(\iota_t\Psi)|_\cals\wedge  e^\calf]_{\rm top}=
[(\iota_m\Psi)|_\cals\wedge  e^\calf]_{\rm top}=0
\eea

Take now $\Sigma\subset M_8$ to be calibrated by $\psi$:
\be
[\psi|_\Sigma\wedge e^{\calf_{\rm mg}}]_{(p)}=\sqrt{\det \hat m}\, \d^p\sigma
\ee
With this prescription on $\Sigma$, one can readily verify that the supersymmety conditions in the form (\ref{calcond2}) are satisfied.
In IIA  with $M_8$ Spin(7) -- see comment after (\ref{eletrback}) -- we have that $\Sigma$ can be a point (D2-brane), or a Cayley four-cycle (D6-brane), or a $M_8$-filling eight-cycle (D8-brane),  or a more complicated configuration including a possible world-volume flux $\calf_{\rm mg}$ too \cite{mmms}.

What remains is to impose  the space-like Bianchi identity $\d\calf_{\rm mg}=0$  and the Gauss law
constraint $\partial_a\hat\Pi^a=0$. The canonical momentum turns out to be given by
\be
\hat\Pi^a=\frac{\mu_{{\rm D}p}}{g_s}\,\sqrt{\det \hat m}\,\hat m^{(ab)}\del_b X
\ee
where $g_s\equiv e^\phi$ is constant and $\hat m^{(ab)}$ is the inverse of $\hat m_{(ab)}$. Now, the Gauss law constraint  (\ref{locgauss}) can be rewritten as
the condition
\be
\mu_{{\rm D}p}\del_a(\sqrt{\det \hat m}\,\hat m^{(ab)}\del_b X)=\sqrt{\det \tilde g}\,\tilde\nabla^2 X=-\mu_{\rm F1}\delta^{(p)}(\del\Sigma_{\rm F1})_{1\ldots p}
\ee
 where $\tilde\nabla_a$ is computed by using the world-volume  metric
 \be
 \tilde g_{ab}=e^{2B}(\hat g_{ab}-\calf_{ac}\hat g^{cd}\calf_{db})
 \ee
 with $e^{2B}=[(\det \hat m)/(\det \hat g|_\Sigma)]^{\frac{2}{2-p}}$.
 Hence, $X$ must be harmonic with respect to the metric $\tilde g$, up to localized sources. If we specialize to a flat internal space $M_8=\mathbb{R}^8$, with the D$p$-brane just spanning $p$ flat directions, we obtain the funnel-shaped solution $X\sim {\ell^{p-1}_s}/r^{p-2}$, which describes a fundamental string ending on the D-brane. This is the BIon of \cite{malda97} and the solution described above is just a generalization of this kind of configuration. 
 
Recall that, in terms of the $(p-1)$ form $\rho_{\rm F1}$ introduced in (\ref{F1diss}), the first term on the r.h.s.\ of (\ref{bED})  can be written as
\be\label{Bionloc}
\int_\Sigma\d^p\sigma\, \hat \Pi^a\omega_a=\mu_{\rm F1}\int_\Sigma \omega\wedge \rho_{\rm F1}
\ee
On the other hand, $\omega|_\Sigma=\d X(\sigma)$. Let us now impose boundary conditions corresponding to $N$ F1 endpoints $\hat\sigma_{(i)}$ on $\Sigma$, $\partial\Sigma_{\rm F1}=\sum^N_{i=1}\hat\sigma_{(i)}$. Then $\d\rho_{\rm F1}=-\sum^N_{i=1}\delta^p(\hat\sigma_{(i)})$ and by integrating by part we obtain that (\ref{Bionloc}) becomes
\be\label{ll}
\mu_{\rm F1} \sum_i\int_\Sigma X(\sigma)\,\delta^p(\hat\sigma_{(i)}) -\mu_{\rm F1}\int_{\del\Sigma} X \rho_{\rm F1}
\ee
If we can approximate $X$ on $\del\Sigma$ to a constant $X_\infty$, by observing that $\int_{\del\Sigma} \rho_{\rm F1}=N$, we see that (\ref{ll}) reduces to
\be
\mu_{\rm F1}\sum^N_{i=1}\Delta X_{(i)}\quad\quad \text{with }\quad\Delta X_{(i)}\equiv X(\hat\sigma_{(i)})-X_\infty
\ee
which is just the energy  of the $N$ strings ending on the D$p$-brane and stretching along the $x$-direction with lengths $\Delta X_{(i)}$.
 
In conclusion, the formula (\ref{bED}) for the energy reduces to
\be
E^{\rm BPS}_{\rm BIon}=\mu_{\rm F1}\sum^N_{i=1}\Delta X_{(i)}+{\mu_{{\rm D}p}}\int_\Sigma e^{-\phi}\psi\wedge e^{\calf_{\rm mg}}
\ee
which, as expected, is just the  sum of the F1 energies and the energy associated with a purely magnetized D-brane.

\subsubsection{Supertubes} 

In the same background $\mathbb{R}^{1,1}\times M_8$ described above, take now a D$(p+2)$-brane of the form $\cals=\mathbb{R}^{1,1}\times \Sigma$. Moreover, we take $\Sigma$ to be of the form $I\times \hat\Sigma$, where 
$I$ is an interval parametrized by the coordinate $\chi$ and $\hat\Sigma=\hat\Sigma(\chi)$ is a one-parameter family of $p$-cycles calibrated by $\psi$. Clearly,  this is possible only if $\hat\Sigma$ is deformable in $M_8$. If there are different possible deformations, we can take an arbitrary one.\footnote{The simplest possibility  \cite{mateos1,mateos2,mateos3,toni} is to take $p=2$, i.e.\ a D2-brane, which implies that $\hat\Sigma$ is just a point.  If for instance $M_8$ is Spin(7) -- see comment after (\ref{eletrback}) -- a point  is obviously calibrated by $\psi$ and then always deformable to any other point.}

We work in partial static gauge by choosing  world-volume coordinates  $\xi^\alpha=(t,x,\chi,\sigma^a)$, $a=1,\ldots, p$. Hence the embedding 
$\hat\Sigma$ is described by 
\be
y^m=Y^m(\chi,\sigma)
\ee
and the calibration condition on $\hat\Sigma$ translates into
\be
\psi|_{\hat\Sigma}=\sqrt{\det\hat g|_{\hat\Sigma}}\,\d^p\sigma
\ee
Take then a world-volume flux of the form
\be
\calf=\d t\wedge\d x+b(\chi)\d x\wedge\d \chi
\ee
The first term on the r.h.s.\ corresponds to the electric field $\cale=\d x$ required by (\ref{stationaryE}).

It is then easy to see that  $ \det \calm =-|b|^2 {\det\hat g|_{\hat\Sigma}}$ while by using (\ref{eletrback}) it is immediate to check that
\bea
&&[(\d t\wedge \Psi)|_\cals\wedge  e^\calf]_{(p+3)}=[(\iota_x \Psi)|_\cals\wedge  e^\calf]_{(p+3)}=b\,\d t\wedge\d x\wedge \d \chi\wedge \psi|_{\hat\Sigma}\cr
&&[(\d x\wedge \Psi)|_\cals\wedge  e^\calf]_{(p+3)}=
[(\d y^m\wedge \Psi)|_\cals\wedge  e^\calf]_{(p+3)}=0\cr &&
[(\iota_t\Psi)|_\cals\wedge  e^\calf]_{(p+3)}=
[(\iota_m\Psi)|_\cals\wedge  e^\calf]_{(p+3)}=0
\eea
Hence, by combining these simple results, one can readily verify that the supersymmetry conditions (\ref{calcond2}) are satisfied
for {\em any} $b(\chi)$ and for any choice {\em any} family of calibrated submanifolds $\hat\Sigma(\chi)$.

What remains is to impose the space-like Bianchi identity and Gauss law. Both are automatically satisfied. The former because $b$ depends just on $\chi$. The latter because $\hat\Pi^a$ has as only non-trivial component $\hat\Pi^x$, which manifestly fulfills  $\partial_x\hat\Pi^x=0$ since $\hat\Pi^x\equiv\hat\Pi^x(\chi,\sigma)$ does not depend on $x$.  

By using $\rho_{\rm F1}$ as defined in (\ref{F1diss}), the total energy is given by
\be
E_{\text{STube}}=\mu_{\rm F1}\int_\Sigma \d x\wedge\rho_{\rm F1}+ \mu_{{\rm D}(p+2)}\int_{\Sigma} \calf_{\rm mg}\wedge e^{-\phi}\psi 
\ee
The first term provides the contribution coming from a number of dissolved F1-strings wrapping a curve $\Sigma_{\rm F1}$ stretched along the $x$-direction. The second term corresponds to the energy associated with a number of dissolved D$p$-branes wrapping an internal cycle $\Sigma_{{\rm D}p}$ belonging to the family $\hat\Sigma(\chi)$, hence calibrated by $\psi$, and appearing as points in $\mathbb{R}^{1,1}$. In other words,  $\rho_{\rm F1}$ and $(2\pi\ell_s)^{-2}\calf_{\rm mg}$ are cohomology classes Poincar\`e dual to $\Sigma_{\rm F1}\subset \Sigma$ and $\Sigma_{{\rm D}p}\subset\Sigma$ and then
\be
E_{\text{STube}}=\mu_{\rm F1}\int_{\Sigma_{\rm F1}} \d x + \mu_{{\rm D}p}\int_{\Sigma_{{\rm D}p}}e^{-\phi}\psi 
\ee
On the other hand, there is no contribution associated with the supporting D$(p+2)$-brane, which appears as `effectively tensionless', consistently with the fact that $\hat\Sigma(\chi)$ is arbitrary -- see \cite{mateos1,mateos2,mateos3,toni} for more discussions.

\section{Conclusion}

In this paper we have obtained a geometric characterization of the supersymmetry conditions for fundamental strings and D-branes.
The analysis is completely general concerning both background and brane configurations. In particular, D-branes can support the most general world-volume flux compatible with supersymmetry. 

The various equations for F1-strings and D-branes look very similar  and come in pairs, with the difference that D-branes `see' a generalized geometry rather than an ordinary one: the covariant supersymmetry conditions are provided by (\ref{F1form2}) and (\ref{calcond2})   [or equivalently (\ref{F1susy}) and (\ref{ggcond})] respectively; the non-covariant minimal BPS conditions are (\ref{F1BPS}) and (\ref{Dcalcond}) respectively; the local BPS bounds saturated by supersymmetric configurations are (\ref{F1bound}) and (\ref{DBPSbound}) [or equivalently (\ref{F1boundm}) and (\ref{DBPSbound2})] respectively.
 
Furthermore, we have shown that the total energy satisfies lower bounds  saturated by its BPS-saturating  values $E_{\rm F1}^{\rm BPS}$ and $E_{{\rm D}p}^{\rm BPS}$ respectively -- see (\ref{BPSenergyF1}) and (\ref{bED}). These are defined by integrals which are invariant under continuous deformations of the brane configuration. Hence $E_{\rm F1}^{\rm BPS}$ and $E_{{\rm D}p}^{\rm BPS}$ can be considered as  `topological' and provide the natural central charge associated with the preserved supersymmetry in presence of F1-strings and D-branes. In particular, our formalism takes into account the effect of the most general world-volume flux on D-branes and can be applied to study D-branes and F1-strings networks extending the approach of \cite{jarahluca}.   

The results obtained here can be useful in several contexts, for instance in the study of black holes or more general black brane configurations.  In particular,  it would be important to study  their implications to  fully coupled bulk-plus-branes systems. With regard to this point, the relation between bulk structures and brane energetics highlighted in this paper should play a key role as it happens for static settings \cite{dimipaul,dwsb}.  Furthermore, it would be interesting to  understand whether one can break the bulk and brane supersymmetry  while preserving (part of) the brane BPS bounds.  This could allow  better control over the fully coupled system in non-supersymmetric settings as well, as for instance it is realized in the non-supersymmetric flux compactifications considered in \cite{dwsb}.

\vspace{1cm}

\centerline{\large\em Acknowledgments}

\vspace{0.5cm}

\noindent I would like to thank  M.~Bianchi, F.~Fucito, P.~Koerber  and specially A.~Tomasiello for useful discussions.  This work is partially supported by the ERC Advanced Grant n.226455 ``Superfields", by the Italian MIUR-PRIN contract 20075ATT78 and by the NATO grant PST.CLG.978785.

\vspace{0.5cm}

%\maketitle  IS IGNORED %%%%%%%%%%%

%\listoftables       % ONLY IN DRAFT MODE
%\listoffigures      % ONLY IN DRAFT MODE

%%%%%%%%%%%%%%%%%%%%%%%%%%%%%%%%%%%%%%%%%%%%%%%%%

\end{document}